%% file: cvb-deletion-channel-journal.tex
\documentclass[journal, twocol]{IEEEtran}

\usepackage[utf8]{inputenc} 
\usepackage[T1]{fontenc}
\usepackage{amsthm}
\usepackage{amsfonts}
\usepackage[ruled,boxed,algonl,linesnumbered,noend]{algorithm2e}
\usepackage[cmex10]{amsmath} 
\usepackage{caption}
\usepackage{cite}
\usepackage{empheq}
\usepackage{graphicx}
\usepackage{ifthen}
\usepackage{mathtools}
\usepackage{setspace}
\usepackage{stmaryrd}
\usepackage{subcaption}
\usepackage{tikz}
\usetikzlibrary{arrows.meta,calc,patterns,positioning,shapes.geometric,arrows,decorations.text}
\usepackage{url}

\graphicspath{{plots/eps/}}

\DeclareMathOperator*{\argmin}{arg\,min}
\DeclareMathOperator*{\argmax}{arg\,max}
\newcommand{\Ed}{{E_{\text{d}}}}
\newcommand{\Ei}{{E_{\text{i}}}}
\newcommand{\Eg}{{E_{\text{g}}}}
\newcommand{\dlbinom}[2]{{\binom{#1}{#2}_{\!\text{d}}}}
\newcommand{\ibinom}[2]{{\binom{#1}{#2}_{\!\text{i}}}}
\newcommand{\gbinom}[2]{{\binom{#1}{#2}_{\!\text{g}}}}

\input{macro.tex}

\begin{document}

\title{Simple Finite-Length Achievability and Converse Bounds for Deletion and Insertion Channels}
\author{
 Ruslan Morozov and Tolga M. Duman \\
 Electrical and Electronics Engineering Department\\
 Bilkent University, Ankara, Turkey\\
 Email: ruslan.morozov@bilkent.edu.tr, duman@ee.bilkent.edu.tr}
\maketitle
{\let\thefootnote\relax\footnote{{This work was funded by the European Union through the ERC Advanced Grant 101054904: TRANCIDS. Views and opinions expressed are, however, those of the authors only and do not necessarily reflect those of the European Union or the European Research Council Executive Agency. Neither the European Union nor the granting authority can be held responsible for them.}}}
{\let\thefootnote\relax\footnote{This work was presented in part at the IEEE International Symposium on Information Theory (ISIT), 2025 \cite{morozov-duman-isit2025}.}}

\begin{abstract}
We develop upper bounds on code size for an independent and identically distributed deletion and insertion channels for a given code length and target frame error probability. 
The bounds are obtained as a variation of a general converse bound, which, though available for any channel, is inefficient and not easily computable without a good reference distribution over the output alphabet. 
We obtain a reference output distribution for a general finite-input finite-output channel and provide a simple formula for the converse bound on the capacity employing this distribution. 
We then evaluate the bound for the deletion channel with a finite block length, and show that the resulting upper bound on the code size is tighter than that for a binary erasure channel, which is the only alternative converse bound for the finite-length setting.
We also provide similar results for the insertion channel.
Furthermore, we present a simple algorithm for computing an achievability bound for a general discrete-input discrete-output channel.
Although the algorithm has exponential complexity, it is useful for comparison purposes.
\end{abstract}

\section{Introduction}

Channels with synchronization errors are encountered in different applications. 
In particular, recent advances in DNA sequencing and the tremendous potential of DNA data storage systems have fueled the study of channels that exhibit insertion and deletion errors. 
Most literature on the capacity of insertion or deletion channels focuses on the asymptotic setting in terms of block lengths, leading to standard information-theoretical results. In addition to such information-theoretic results on synchronization error channels, it is crucial to study the scenarios with finite-block-length inputs and outputs. 
For instance, focusing on DNA data storage systems, while some sequencing technologies, e.g., nanopore, can support DNA strands up to a few hundred thousand nucleotides \cite{deamer16three}, many other technologies synthesize and read only DNA strands of a few hundred nucleotides  \cite{church12next, goldman13towards, antkowiak20low, mao18models, hamoum23channel,welter24end}.
With this motivation, in this paper, we study synchronization error channels with finite block lengths rather than the asymptotic setting that yields standard information-theoretical results. We focus on the binary alphabet case for ease of exposition. Specifically, we consider the case of binary deletion channels (BDCs) and binary insertion channels (BICs), i.e., channels with independent and identically distributed (i.i.d.) deletions (insertions) with non-vanishing deletion (insertion) probability, and obtain finite-length converse bounds.
We also provide a straightforward greedy algorithm for computing an achievability bound for a general discrete-input discrete-output channel, which is mainly used for comparison purposes. 

Consider a channel $W$. 
Let $\bfM(W,\ve)$ be the maximum code size for the channel $W$ and target decoding error probability $\ve$. 
A lower bound on $\bfM(W,\ve)$ is called an \emph{achievability bound (AVB)}.
It gives information about the existence of at least \emph{one} coding scheme, which has a given code size and error probability not greater than $\ve$. 
An upper bound on $\bfM(W,\ve)$ is called a \emph{converse bound (CVB)}, and it is usually harder to obtain, since it is an upper bound on the rates of \emph{all} possible coding/decoding schemes over $W$ with the decoding error probability not greater than $\ve$.
A straightforward approach would require a brute-force search over all possible codes and the analysis of the error probability of maximum a posteriori probability (MAP) decoding.
This is feasible only for toy examples (e.g., for input block lengths around 5 bits); hence, we need to take a different approach.

As a practical benchmark, in addition to the channel capacity, which assumes an asymptotic setting in the block length, the normal approximations derived in \cite{polyanskiy2010channel} are also employed (e.g., see \cite{piano2020approaching}). However, neither the channel capacity nor the normal approximations are strict bounds on the performance of channel codes for a fixed code length and error probability. Note that one can obtain a true bound using the normal approximation results if the asymptotic $O$-terms are also bounded, which could be possible using different techniques. On the other hand, we generally do not have a simple bound on these constants, and the results are only approximate. Note also that the analysis results based on the channel capacity and the normal approximation become more accurate for memoryless (product-like) channels as codeword length increases. 
On the contrary, our focus here is the BDC/BIC with input lengths up to a few hundred bits, hence the above benchmarks are not necessarily tight.

A classical finite-length CVB is the sphere-packing bound \cite{shannon93lower}, Theorem~2.
Finite-length converse bounds for BEC and BSC are available in \cite{polyanskiy2010channel}. Specifically, in \cite{polyanskiy2010channel}, Theorem~28, there is a CVB for channels that have a certain symmetry. 
Namely, the hypothesis test errors between conditional and unconditional output distributions should be connected in the same way for all inputs. 
This is true for the additive white Gaussian noise (AWGN) channel, and the above bound is computed numerically in \cite{erseghe15evaluation}.
In~\cite{kalai10tight}, Theorem~II.1, there is a finite-length CVB for the deletion channel.
Although it is asymptotically optimal, meaning that, with $N\to\infty$, and then with $\delta\to 0$, the bound approaches the lower bound on the capacity, this bound is not useful for non-asymptotic values of $\delta$ and $N$. Also, in Theorem~II.2, the authors provide a finite-length upper bound on the capacity of the deletion channel with a fixed number of deletions (BDCFND). This bound, however, is not tight for block lengths of interest, and is used only as an intermediate step to obtain a channel capacity expansion in the asymptotic regime of near-zero deletion probabilities. 
The same trick is used in~\cite{li14capacity}, where, together with a brief summary on results of upper bounds on the capacity of the deletion channel, an upper bound on the capacity of the BDCFND is presented. 
A thorough overview of the capacity results for the BDC is presented in~\cite{cheraghchi21overview}.

Technically, the performance of any existing coding scheme can serve as an AVB. 
Another approach is to assume a decoder, which outputs $x$, if the mutual information density for $x$ is larger than some threshold.
Such an approach is employed in the dependency testing (DT) bound \cite{polyanskiy2010channel}.
It is an AVB, available for a general channel.
In the context of insertion/deletion channels, however, obtaining the distribution of mutual information density is a hard problem.
In \cite{maarouf23finite}, this problem is addressed by resorting to Monte-Carlo simulations, and achievability results with finite block lengths are estimated. In addition, estimates of achievable rates are presented in \cite{mcbain24information} for a DNA nanopore sequencing channel, modeled as Markov duplications plus AWGN.
For a realistic DNA channel model, which includes sampling multiple times, both achievability and converse finite-length bounds are presented in \cite{weinberger22dna}. 

In \cite{tan15asymptotic}, Section~4.1.2, a CVB, called symbol-wise converse bound, for a general channel is presented (see also Th.~4 in \cite{verdu94general}). 
It is based on hypothesis testing between the conditional distribution and some reference output distribution.
However, in the case of BDC, a randomly picked reference distribution leads to a very loose bound.
Moreover, the complexity of computing the bound for a deletion channel with hundreds of input bits is infeasible.

Ref. \cite{fertonani11upper} presents the methodology to obtain the tightest upper bound on the capacity of deletion channels for a large range of deletion probabilities \footnote{See also \cite{rubinstein2023} which uses the same technique, but with longer input block lengths to improve the bounds via more refined numerical computations.}. The paper employs a trick to analyze a deletion channel with $N$ input bits, namely, to move to $n$ independent uses of a deletion channel with fewer ($m$) bits, such that $N\approx mn$, by giving a side information to the receiver.
In this paper, we employ this trick as well: we simplify and compute the symbol-wise converse bound for $n$ independent uses of the deletion channel with $m$ input bits. 
Since the original deletion channel with $nm$ input bits is degraded with respect to its information-aided memoryless version, the computed CVB automatically serves as a CVB for the original deletion channel. 
For the deletion channel with side information, we provide a family of good reference distributions and obtain the symbol-wise converse bound for these distributions in closed form.
Moreover, to obtain a list of good candidate distributions, we split the output alphabet into subsets of equal length, referred to as \emph{layers}.
We construct useful output distributions, which cover only a subset of layers, and obtain CVB for deletion and insertion channels. 
One can see a parallel between the layer-oriented distributions and the BDCFND, described above.

The paper is organized as follows. In Section~\ref{s:back}, we recall the symbol-wise converse bound. In Section~\ref{s:mocvb}, we introduce a max-oriented output distribution, which can be used for the implementation of the symbol-wise converse bound. In Section~\ref{s:locvb}, we present a generalization of a max-oriented distribution, namely, a family of layer-oriented distributions.
We show how to compute the symbol-wise converse bound for a layer-oriented distribution and minimize the bound over the provided family of distributions.
In Section~\ref{s:other}, we recall the normal approximation for a memoryless channel.
We also provide a simple (but high-complexity) algorithm for computing an achievability bound for a general discrete channel.
We compute both of them for the deletion (insertion) channel for different block lengths.
In Section~\ref{s:num}, we compare the proposed bounds for the deletion (insertion) channel with optimal codes, as well as with other bounds and approximations. We observe that the newly developed converse bounds are superior to the alternatives; however, they are not very close to the achievability results, hence there is room for further improvement.

\section{Background}
\label{s:back}

\subsection{Notation and Channel Models}
Vectors and subvectors are denoted as $v_a^b=(v_a,\dots,v_b)$.
A subvector of $v$ with indices from set $\mS=\set{s_1,\dots,s_t}$, $s_{i}<s_{i+1}$ is denoted by $v_{\mS}=(v_{s_1}, \dots, v_{s_t})$.

Denote by $W:\mX\tosq\mY$ a discrete channel\footnote{We use $\tosq$ and not $\to$ to distinguish a channel from a mapping from $\mX$ to $\mY$.} with finite set of input symbols $\mX$ and finite set of output symbols $\mY$, and the conditional probabilities $W(y|x)$.
Denote by $\bfM(W, \ve)$ the maximum size $|\mC|$ of a code $\mC\subseteq\mX$, such that there exists a decoding function $f:\mY\to\mC$, and the average error probability is not higher than $\ve$ for uniform input distribution over $\mC$:
\begin{align}
\frac{1}{|\mC|}\cdot \sum_{c\in\mC}\Pr_{W(Y|c)}\brsq{f(Y)\neq c} \leq \ve.
\label{eq:decf}
\end{align}

Denote the binary i.i.d. deletion channel with $m$ input bits and deletion probability $\delta$ by $D_m^{(\delta)}$.
The deletion channel removes each input bit independently with probability $\delta$, or transmits it perfectly with probability $1-\delta$.
The transition probabilities of $D_m^{(\delta)}:\bF_2^m\tosq \bF_2^{\leq m} = \cup_{w=0}^m \bF_2^w$, where $\bF_2=\set{0,1}$,  are defined by
\begin{align}
D^{(\delta)}_m(y_1^w|x_1^m)=\dlbinom{x_1^m}{y_1^w}\cdot \delta^{m-w} \cdot (1-\delta)^w,
\label{eq:bdc}
\end{align}
where $w\in\set{0,1,\dots,m}$ is the output length, and for two binary strings $x_1^m$ and $y_1^w$, the \emph{d-embedding number} $\dlbinom{x_1^m}{y_1^w}$ is defined as the number of subsequences of $x$ which are equal to $y$:
\begin{align}
\dlbinom{x_1^m}{y_1^w}=\brabs{\set{\mS\subseteq \set{1,\dots,m} \mid x_\mS=y_1^w}}.
\label{eq:embdef}
\end{align}

We also consider two different models for insertions. 
In the first model, after each input symbol, one symbol is inserted with probability $\iota$. The inserted bits are independent of the transmitted bits and of each other, and they are selected uniformly on $\{0,1\}$. 
Such binary insertion channel with $m$ input bits and insertion probability $\iota$ is denoted by $I_m^{(\iota)}$. 
The transition probabilities of $I_m^{(\iota)}:\bF_2^m\tosq \cup_{w=m}^{2m} \bF_2^w$ are defined by
\begin{align}
I^{(\iota)}_m(y_1^w|x_1^m)=\ibinom{y_1^w}{x_1^m} \cdot \frac{\iota^{w-m} \cdot (1-\iota)^{2m-w}}{2^w},
\label{eq:bic}
\end{align}
where $w\in\set{m,m+1,\dots,2m}$ is the output length, and for two binary strings $x_1^m$ and $y_1^w$, the \emph{i-embedding number} $\ibinom{y_1^w}{x_1^m}$ is the number of ways of inserting $(w-m)$ symbols (maximum $1$ symbol after each symbol) into $x_1^m$ to obtain $y_1^w$.
Equivalently, it is the number of sets $\mS\subseteq \set{2,\dots, w}$, $|\mS|=w-m$, which do not include a pair of adjacent numbers, such that $x$ is a subsequence of $y$ with indices \emph{not} from $\mS$: $y_{\set{1,\dots,w}\setminus\mS}=x_1^m$.
The difference between $\dlbinom{y}{x}$ and $\ibinom{y}{x}$ is that in $\ibinom{y}{x}$ we only count those deletion patterns in $y$, which do not include adjacent symbols, and we also cannot delete the first symbol, so $\ibinom{y}{x}\leq \dlbinom{y}{x}$.

In the second insertion model, referred to as \emph{Gallager's insertion channel}, each input bit is either transmitted perfectly, or the input bit is deleted, and two random bits are transmitted instead of it.
Denote such channel with $m$ input bits and insertion probability $\gamma$ by $G_m^{(\gamma)}$.
Then, the transition probabilities of $G_m^{(\gamma)}:\bF_2^m\tosq \cup_{w=m}^{2m} \bF_2^w$ are defined by
\begin{align}
G^{(\gamma)}_m(y_1^w|x_1^m)=\gbinom{y_1^w}{x_1^m} \cdot \frac{\gamma^{w-m} \cdot (1-\gamma)^{2m-w}}{4^w},
\label{eq:gc}
\end{align}
where the \emph{g-embedding number} $\gbinom{y_1^w}{x_1^m}$ is equal to the number of ways to obtain $y_1^w$ from $x_1^m$ by the described above insertions.
More formally, we can define the g-embedding number as the number of pairs $(\mS,\mT)$, such that $y_\mS=x_\mT$, where sets $\mS=\set{s_1,\cdots,s_u}$, $1\leq s_1 < \dots <s_u\leq w$, and $\mT=\set{t_1,\cdots,t_u}$, $1\leq t_1 < \dots <t_u\leq m$ satisfy $s_{i+1}-s_i-1=2(t_{i+1}-t_i-1)$ for $i=0,\dots,u$, assuming $s_0=t_0=0$ and $s_{u+1}=w+1$, $t_{u+1}=m+1$.

\subsection{The Symbol-Wise Converse Bound}

The symbol-wise converse bound in  \cite[Prop 4.4]{tan15asymptotic} states that for any channel $W:\mX\tosq\mY$ and $\eta\in(0,1-\ve)$:
\begin{align}
\log_2\bfM(W,\ve) \leq \inf_{Q} \sup_{x\in\mX} \bfD_{\ve+\eta}(W(\cdot|x)||Q)+\log_2\frac{1}{\eta},
\label{eq:cvborig}
\end{align}
where $\bfD_{\vp}(P||Q)$ denotes the $\vp$-information spectrum divergence between distributions $P$ and $Q$, defined as
\begin{align}
\bfD_{\vp}(P||Q)= \sup\set{r \bigg| \Pr_{P(Z)}\brsq{\log_2\frac{P(Z)}{Q(Z)}\leq r}\leq \vp}.
\label{eq:dse}
\end{align}
By letting $\vp=\ve+\eta$ and exponentiating of both sides, one can rewrite \eqref{eq:cvborig} as
\begin{align}
&\bfM(W,\ve) \leq \obfM_{\text{SW}}(W,\ve,Q,\vp)
\nonumber \\&
\ \ \ \defeq \frac{1}{\vp - \ve} \cdot \max_{x\in\mX} \sup \set{\rho\bigg| \Pr_{W(Y|x)}\brsq{\frac{W(Y|x)}{Q(Y)}\leq \rho} \leq \vp}
\label{eq:cvb}
\end{align}
for $\vp > \ve$.
Observe that we can minimize $\obfM_{\text{SW}}(W,\ve,Q,\vp)$ over $\vp$ and $Q$ without violating the bound.

Computing \eqref{eq:cvb} for a deletion channel $D_m^{(\delta)}$ for large $m$ is infeasible since there is no sub-exponential algorithm for computing all channel transition probabilities.
For example, for $m$ as small as $50$, a deletion channel is given by $2^{50}$ conditional distributions over $2^{51}-1$ possible outputs.
Moreover, it is infeasible to minimize \eqref{eq:cvb} over all distributions $Q$.

Picking a good distribution $Q$ is also a non-trivial problem. In this paper, we simplify the bound and propose some distributions $Q$ for which the bound is easy to compute.
Also, the proposed distributions below lead to good bounds in the sense that we could not improve them by selecting different $Q$ through random search and optimization via genetic algorithms.

\subsection{The BEC Bound}

The Polyanskiy-Poor-Verd\'u converse bound \cite{polyanskiy2010channel} (Theorem 38) states that for any code of size $M$ over BEC $E_n^{(\delta)}$ with $n$ input bits and erasure probability $\delta$, the decoding error probability $\ve$ is lower-bounded by
\begin{align}
\ve\geq \sum_{l=\floor{n-\log_2 M}+1}^n \binom{n}{l} \delta^l (1-\delta)^{n-l} \br{1-\frac{2^{n-l}}{M}}.
\label{eq:bec}
\end{align}
The BDC $D^{(\delta)}_n$ and the Gallager's insertion channel $G_n^{(\delta)}$ are degraded with respect to $E_n^{(\delta)}$ as a BEC channel is obtained by revealing the positions of deletions and insertions as side information \cite{fertonani10novel, fertonani11upper}. Thus, the error probability for the code of size $M$ over the BDC is also lower-bounded by \eqref{eq:bec}.
Note that using binary search, one can translate this bound to an upper bound on $\bfM(D^{(\delta)}_n,\ve)$ and $\bfM(G^{(\delta)}_n,\ve)$ as well.
We will call the latter bound \emph{the BEC bound} for the BDC or for the Gallager's insertion channel.

\section{The Max-Oriented Converse Bound for a General Discrete Channel}
\label{s:mocvb}

In the general symbol-wise converse bound \eqref{eq:cvb}, one of the degrees of freedom is the reference distribution $Q$.
For each such $Q$, for a fixed probability $\vp$, the bound is equal to the maximum (over $x$) of $\vp$-quantile $\rho$ of $W(Y|x)/Q(Y)$.
The smaller the quantile, the tighter the bound.
The roles of all variables in this bound are dual to each other: if we pick a larger probability $\vp$, then the value of $1/(\vp-\ve)$ is smaller, but the maximum $\vp$-quantile is larger.

Below we provide a special case of \eqref{eq:cvb}, called \emph{the maximum-oriented converse bound}.
This bound can be easily computed for any channel if all the input-output (conditional) probabilities are given.
First, for a channel $W$, define
\begin{align}
\tau(W)&\;\defeq\;\sum_y \max_x W(y|x).
\label{eq:tau}
\end{align}
Then, the following holds.

\begin{proposition}
\label{p:mocvb}
The maximum code size which allows transmission over channel $W$ with error probability $\ve$ is upper-bounded by
\begin{align}
\bfM(W, \ve) \leq \frac{\tau(W)}{1-\ve}.
\label{eq:p1w}
\end{align}
\end{proposition}

\begin{proof}
Consider the case of $\vp=1$ in \eqref{eq:cvb}.
The $1$-quantile is just the maximum possible value of $W(y|x)/Q(y)$:
\begin{align}
&\obfM_{\text{SW}}(W,\ve,Q,1) 
\nonumber\\ & \ \ =
 \frac{1}{1 - \ve} \cdot \max_x\inf \set{\rho\bigg| \Pr_{W(Y|x)}\brsq{\frac{W(Y|x)}{Q(Y)}\leq  \rho}= 1}
\nonumber\\ &
\ \ = \frac{1}{1 - \ve} \cdot \max_y \frac{\max_x W(y|x)}{Q(y)}.
\label{eq:mvp1}
\end{align}
Due to the constraint $\sum_y Q(y)=1$, the minimum of \eqref{eq:mvp1} over $Q$ is achieved, when $\max_x W(y|x)/Q(y)$ is the same for all probable\footnote{If $W(y|x)=0$, then $y$ does not contribute to $\Pr[..]$ in \eqref{eq:mvp1}.} values of $y$, i.e., when $Q(y)$ is proportional to $\max_x W(y|x)$.
Denote such distribution by $\tQ_W$: 
{\allowdisplaybreaks
\begin{align}
\tQ_W(y)&\;\defeq\;\frac{\max_x W(y|x)}{\tau(W)}.
\label{eq:tQ}
\end{align}
}%
Note that it is optimal for $\vp=1$, i.e., 
\begin{align}
\obfM_\text{SW}(W,\ve,\tQ_W,1)=\inf_Q\obfM_\text{SW}(W,\ve,Q,1).
\label{eq:tQopt}
\end{align}
Substituting distribution $\tQ_W$ in \eqref{eq:mvp1}, one obtains
\begin{align}
\bfM(W,\ve)\leq \obfM_\text{SW}(W,\ve,\tQ_W,1)=\frac{\tau(W)}{1-\ve}.
\label{eq:cvbs}
\end{align}
\end{proof}

We denote the right-hand-side of \eqref{eq:cvbs} by 
\begin{align}
\obfM_\textsc{MO}(W,\ve)\ \defeq\ \obfM_\text{SW}(W,\ve,\tQ_W,1) = \frac{\tau(W)}{1-\ve}
\end{align}
and we call it \emph{the max-oriented converse bound (MO-CVB)} for $W$.
We also call $\tQ_W$ \emph{the max-oriented distribution}. 

\begin{proposition}
\label{p:taun}    
For a channel $W^n$, the value of $\tau$ is given by
\begin{align}
\tau(W^n)&=\tau(W)^n.
\label{eq:tauwn}
\end{align}
\end{proposition}
\begin{proof}
The maximization over all $x_1^{n}$ of $W^n(y_1^n|x_1^n)$ is independent for each $x_i$, hence we can write
\begin{align}
\tau(W^n)&=\sum_{y_1^{n}\in\mY^n}\max_{x_1^n \in \mX^n} W^n(y_1^n|x_1^n)
\nonumber\\&
=\prod_{i=1}^n \sum_{y\in	\mY} \max_{x \in \mX} W(y|x)=\tau(W)^n.
\label{eq:tauwnp}
\end{align}
\end{proof}

Combining Propositions~\ref{p:mocvb} and \ref{p:taun}, we obtain a max-oriented converse bound for multiple independent uses of a channel $W$:
\begin{align}
\obfM_\text{MO}(W^n,\ve) = \frac{\tau(W)^n}{1-\ve}.
\label{eq:Mn}
\end{align}

Denote the corresponding code rate (in bits/symbol) by 
\begin{align}
\overline{r}(n,\ve)=\frac{\log_2 \obfM_\text{MO}(W^n,\ve)}{n\log_2 |\mX|}.
\label{eq:rWn}
\end{align}
The dynamics of $\overline{r}(n,\ve)$ for increasing $n$ is as follows:
{\allowdisplaybreaks
\begin{align}
&\frac{\overline{r}(n,\ve)}{\overline{r}(n-1,\ve)}
=\frac{n-1}{n}\cdot \frac{n\log_2\tau(W)-\log_2(1-\ve)}{(n-1)\log_2\tau(W)-\log_2(1-\ve)}
\nonumber \\&
= \frac{\log_2\tau(W)-\frac{\log_2(1-\ve)}{n}}{\log_2\tau(W)-\frac{\log_2(1-\ve)}{n-1}}<1
\end{align}
}%
which means $\overline{r}(n,\ve) < \overline{r}(n-1,\ve)$ for $0<\ve<1$.
Thus, for a larger $n$, the upper bound on the code rate is lower.
When the length goes to infinity, we can obtain the limit
\begin{align}
\overline{r}(\infty,\ve) & =\lim_{n\to\infty} \frac{\log_2\oM_\vp(W^n)}{n\log_2 |\mX| }
\nonumber \\ & =
\lim_{n\to\infty} \frac{n\log_2\tau(W)-\log_2(1-\ve)}{n\log_2|\mX|} 
\nonumber \\ & =\log_{|\mX|}\tau(W),
\label{eq:rhoinf}
\end{align}
which can be considered as an upper bound on the channel capacity\footnote{For the channels considered in this paper, this bound is looser than the existing ones.}.

\section{The Layer-Oriented Converse Bound}
\label{s:locvb}
\subsection{The Layer-Oriented Bound for the General Channel}

In this section, we generalize the MO-CVB (Proposition~\ref{p:mocvb}), developed in the previous section.
If, for a given channel, there exist \emph{layers}, i.e., independent subsets of output symbols, for which a local max-oriented reference output distribution can be constructed, then we obtain a layer-oriented converse bound, which in general is tighter than the MO-CVB.

Consider a channel $W:\mX \tosq \mY$.
Let $\mL \subseteq \mY$ be a subset of outputs which is equiprobable for any input:
\begin{align}
\forall x\in\mX: \sum_{y\in\mL} W(y|x)=p(W,\mL).
\label{eq:mZi}    
\end{align}
We will call such a subset \emph{a layer}.
Note that the union of two layers is also a layer.
The trivial layers are $\mL=\mY$ and $\mL=\emptyset$.

\begin{proposition}
For any layer $\mL$,   
\begin{align}
\bfM(W,\ve) \leq \frac{\tau(W,\mL)}{p(W,\mL)-\ve}.
\label{eq:locvb}    
\end{align}
\end{proposition}
\begin{proof}
Since a layer is equiprobable for all inputs, its probability contributes to the overall quantile order ($\rho$ in \eqref{eq:cvb}).
One can construct a local maximum-oriented distribution over a layer $\mL$ and obtain a \emph{layer-oriented distribution}.
That is,
\begin{align}
\tQ_{W,\mL}(y)  \ &\defeq \
\begin{cases}
\displaystyle\frac{\max_x W(y|x)}{\tau(W,\mL)}, & y\in \mL   \\
0, & y\notin \mL
\end{cases}
\label{eq:loq}
\\
\tau(W,\mL) \ &\defeq \ \sum_{y \in \mL} \max_x W(y|x).
\end{align}
Applying the distribution $\tQ_{W,\mL}$ to the converse bound \eqref{eq:cvb}, we obtain \emph{the layer-oriented converse bound} (LO-CVB) $\obfM_\text{LO}(W,\ve,\mL)=\obfM(W,\ve,\tQ_{W,\mL},p(W,\mL))$, which is given by
\begin{align}
\bfM(W,\ve) \leq \obfM_\text{LO}(W,\ve,\mL) = \frac{\tau(W,\mL)}{p(W,\mL)-\ve}.
\label{eq:locvb}    
\end{align}
Note that \eqref{eq:locvb} only holds if the denominator is positive.
\end{proof}

For the memoryless channel $W^n$ which is equivalent to $n$ independent copies of $W$, with the same approach as in Proposition~\ref{p:taun}, we can obtain
\begin{align}
\bfM(W^n,\ve)\leq \obfM_\text{LO}(W^n,\ve,\mL^n) = \frac{\tau(W,\mL)^n}{p(W,\mL)^n-\ve},
\label{eq:locvbn}    
\end{align}
if $p(W,\mL)^n>\ve$.

\subsection{A Simpler Derivation of LO-CVB from Scratch}
\label{s:mos}

A simpler derivation of the LO-CVB (including the MO-CVB as a special case) can be obtained without using the symbol-wise bound.
Consider a code $\mC$ of size $M$.
The input message is distributed uniformly over $\mC$, so, $\Pr\set{x}=\frac{1}{M}$.
The optimal decoder, upon receiving $y$, outputs some $x\in\arg\max_{x\in\mC}W(y|x)$.
Assuming that the decoding is always correct when $y\notin\mL$, and is MAP when $y\in\mL$, we can lower-bound the probability of decoding error by
\begin{align}
\ve &\geq  \Pr\set{y\in\mL}-\Pr\set{y\in\mL \wedge \text{correct dec.}} \nonumber\\&
= p(W,\mL)-\sum_{y \in\mL} \frac{1}{M} \cdot \max_{x\in\mC} W(y|x) 
= p(W,\mL)-\frac{\tau(W,\mL)}{M},
\label{eq:locvbd}
\end{align}
which imples the LO-CVB
\begin{align}
\bfM(W,\ve) \leq \frac{\tau(W,\mL)}{p(W,\mL)-\ve}.
\label{eq:locvbs}
\end{align}

\subsection{The Layer-Oriented Bound for the Deletion Channel}

Consider the deletion channel $D^{(\delta)}_m:\bF_2^m\tosq \bF_2^{\leq m}$, defined in \eqref{eq:bdc}.
Since the output length does not depend on the input, we can employ the sets of the form $\bF_2^w$ for each $w$ as the layers of $D^{(\delta)}_m$. 
The values of $\tau$ and $p$ for these layers are given by
\begin{align}
\tau(D^{(\delta)}_m, \bF_2^w)&= \Ed(m,w) \cdot \delta^{m-w}(1-\delta)^w
\label{eq:taulD}
\\
p(D^{(\delta)}_m, \bF_2^w)&=\binom{m}{w}\cdot \delta^{m-w}(1-\delta)^w,
\label{eq:pD}
\end{align}
where $\Ed(m,w)$ is the sum of the maximal d-embedding numbers, defined in \eqref{eq:embdef}:
\begin{align}
\Ed(m,w)=\sum_{y_1^w} \max_{x_1^m}\dlbinom{x_1^m}{y_1^w}.
\label{eq:Emw}
\end{align}
For example,
\begin{align}
\Ed(5,2)&=32,\ \Ed(5,3)=52,\ \Ed(5,4)=54
\nonumber \\
\Ed(m,0)&=1,\ \Ed(m,1)=2m,\ \Ed(m,m)=2^m.
\label{eq:Eex}
\end{align}

Any union of layers is also a layer, so one can obtain a layer by the union of any subset of $\set{\set{}, \bF_2,\bF_2^2,\dots, \bF_2^m}$.
Moreover, denote by $n$ the number of independent uses of channel $D_m^{(\delta)}$.
The idea is to provide, instead of a converse bound for the channel $D^{(\delta)}_{mn}$, a converse bound for the channel $(D^{(\delta)}_{m})^n$. This approach works since the channel $(D^{(\delta)}_{m})^n$ is equivalent to the channel $D^{(\delta)}_{mn}$ with the additional side information on the boundaries of received blocks corresponding to each block of $m$ input symbols at the receiver. Denote the LO-CVB for a given subset $\Lambda$ of output lengths by

\begin{gather}
L(n,m,\ve,\Lambda)\defeq
\frac{\br{\displaystyle\sum_{w\in\Lambda}\tau\br{D^{(\delta)}_m,\bF_2^w}}^{\!n}}{\br{\displaystyle\sum_{w\in\Lambda} p\br{D^{(\delta)}_m,\bF_2^w}}^{\!n}-\ve},
\label{eq:L} 
\end{gather}
where we assume $L(n,m,\ve,\Lambda)=+\infty$ if the denominator is non-positive.
The resulting LO-CVB for the deletion channel $D_{mn}^{(\delta)}$ is defined as
\begin{align}
\bfM(D_{mn}^{(\delta)},\ve)\leq \bfM((D_{m}^{(\delta)})^n,\ve)\leq \min_{\Lambda\subseteq\set{0,1,\dots, m}} L(n,m,\ve,\Lambda).
\label{eq:locvbdel}
\end{align}
Observe that the MO-CVB is a special case of LO-CVB, when $\Lambda=\set{0,1,\dots,m}$.
If $\Lambda \neq\set{0,1,\dots,m}$, then $\sum_{w\in\Lambda}p\br{D_m,\bF_2^w}<1$, and the first term in the denominator of \eqref{eq:L} decreases exponentially, meaning that, for sufficiently large $n$, the first term becomes less than $\ve$, which leads to the invalid case.
This means that for sufficiently large $n$ we can only use $\Lambda=\set{0,1,\dots,m}$, and the LO-CVB becomes the MO-CVB. 

Note that computing $\Ed(m,w)$ is the only non-trivial problem in computing the LO-CVB for the deletion channel.
The straightforward algorithm is to run over all $x_1^m$ and for each of them perform all deletion patterns of weight $(m-w)$.
The complexity of such approach is $O\br{ m\cdot \binom{m}{w} \cdot 2^m}$.
By exploiting reverse and inverse symmetries, one can reduce the complexity by approximately $4$ times.
The complexity of computing the complete set of $\Ed(m,w)$ for all $w=0,\dots,m$ is $O(m\cdot 2^{2m})$.
After the complete set is computed, we can obtain the LO-CVB by \eqref{eq:locvbdel}\footnote{Another optimization would be to run over only those $\Lambda$, which constitute contiguous segments of output lengths. 
Such optimization results in negligible loosening of the bound. 
But the complexity of computing $\Ed(m,w)$ is much higher than running over all sets of lengths, so this is not crucial.\label{ft:opt}}.

For a fixed $N=mn$, the LO-CVB becomes tighter with larger $m$, since we provide less side information to the receiver.
Due to high complexity, the complete sets of $\Ed(m,0), \Ed(m,1), \dots, \Ed(m,m)$ can only be obtained for small $m$.
For larger values of $m$, however, we still can compute $\Ed(m,w)$ for $w$ close to either $0$ or $m$, since in these cases the binomial coefficient $\binom{m}{w}$ is smaller.
On the other hand, for a valid bound, we must keep the denominator of \eqref{eq:L} positive.
Thus, the partial computation of $\Ed(m,w)$ allows us to optimize the LO-CVB when the deletion probability is very small or very large, and the length of the output is close to $0$ or to $m$ with high probability.
For the incomplete set of values of $\Ed(m,w)$, we minimize \eqref{eq:locvbdel} over subsets $\Lambda$ of all lengths $w$, for which $\Ed(m,w)$ is available.

\subsection{The Layer-Oriented Bound for the Insertion Channel}

The binary insertion channel $I_m^{(\iota)}$ is defined in \eqref{eq:bic}.
Similarly to the case of the deletion channel, the output length does not depend on the input, so again the sets $\bF_2^w$, where $w=m\dots 2m$, are the layers of $I_m^{(\iota)}$.
The values of $\tau$ and $p$ for these layers are given by
\begin{align}
\tau(I^{(\iota)}_m, \bF_2^w)&= \Ei(m,w) \cdot \iota^{w-m}(1-\iota)^{2m-w}
\label{eq:tauI}
\\
p(I^{(\iota)}_m, \bF_2^w)&=\binom{m}{w-m}\cdot \iota^{w-m}(1-\iota)^{2m-w},
\label{eq:pI}
\end{align}
where $\Ei(m,w)$ is the sum of maximal i-embedding numbers 
\begin{align}
\Eg(m,w)=\sum_{y_1^w} \max_{x_1^m} \ibinom{y_1^w}{x_1^m},
\label{eq:E1wm}
\end{align}
and the symbol $\ibinom{y_1^w}{x_1^m}$ is defined in Section~\ref{s:back}.
For example, 
\begin{gather}
\Ei(2,3)=12,\ \Ei(m,m)=2^m,\ \Ei(m,2m)=2^{2m}.
\label{eq:e1example}
\end{gather}

Computation of $\Ei(m,w)$ requires going over all $x_1^m\in \bF^m$, and all the $\binom{m}{w-m}$ possible sets of the inserted positions.
Furthermore, for each such $x_1^m$ and the set of the inserted positions, we run over all $2^w$ possible values of the inserted bits.
For each obtained pair $(x_1^m,y_1^w)$, we count the number of possible insertion configurations (i.e., the set of the inserted positions together with the value of the inserted bits).
For each obtained $y_1^w$, we are storing the current maximum value of $\ibinom{y_1^w}{x_1^m}$ over all $x_1^m$ that have already been processed.
While running over all $x_1^m$, we update the maximum of the i-embedding numbers when we find $x_1^m$, for which $\ibinom{y_1^w}{x_1^m}$ is larger than the current maximum.
The detailed algorithm is given in Algorithm~\ref{alg:e12}, which computes $\Ei(m,w)$ when the input $t=\text{i}$.

\begin{algorithm}
\caption{Computation of $\Ei(m,w)$ or $\Eg(m,w)$.}
\label{alg:e12}
\DontPrintSemicolon
\KwIn{$m, w\in\bN$, $0<m\leq w \leq 2m$, and $t\in\set{\text{i},\text{g}}$}
\KwOut{$E_t(m,w)$}
$P \gets$ the set of all $m$-bit integers which have $(w-m)$ ones in their binary representation \\
$S\gets 2^{tw}$ \\
$\var{E} \gets $ array of zeroes of size $S$ \\
\For{$x\gets 0,\dots, 2^m-1$}{
    $\var{e} \gets $ an empty priority queue with integer keys and integer values \\
    \For{$p\in P$}{
        \For{$b\gets 0\dots S-1$}{
            $y \gets 0$ \\
            \For{$i\gets 0\dots m-1$}{
                \eIf(// \textit{regular insertion:}){$t=$ {\normalfont i}}{
                    $y\gets 2y+(x \bmod 2)$\\
                    $x \gets \floor{x/2}$ \\
                    \If{$(p\bmod 2)=1$}{
                        $y \gets 2y + (b\bmod 2)$ \\
                        $b \gets \floor{b/2}$ \\
                    }
                    $p \gets \floor{p/2}$ \\
                }(// \textit{Gallager insertion:}){
                    \eIf{$(p\bmod 2)=1$}{
                        $y \gets 4y + (b\bmod 4)$ \\
                        $b \gets \floor{b/4}$ \\
                    }{
                        $y\gets 2y+(x \bmod 2)$
                    }
                    $x \gets \floor{x/2}$ \\
                    $p \gets \floor{p/2}$ \\
                }
            }
            \lIf{$y\notin$ keys of \var{e}}{$\var{e}[y] \gets 0$}
            $\var{e}[y] \gets \var{e}[y] + 1$            
        }    
    }
    \For{$y\in$ keys of \var{e}}{
        $\var{E}[y] \gets \max\set{\var{E}[y], \var{e}[y]}$        
    }    
}
\textbf{return} $\displaystyle\sum_{y=0}^{2^w-1}\var{E}[y]$
\end{algorithm}

\subsection{The Layer-Oriented Bound for the Gallager Insertion Channel}

As in the previous cases, in the Gallager's insertion channel, defined in \eqref{eq:gc}, the output length also does not depend on the input, so again the sets $\bF_2^w$ are the layers of $G^{(\gamma)}_m$.
The values of $\tau$ and $p$ for these layers are given by
\begin{align}
\tau(G^{(\gamma)}_m, \bF_2^w)&= \Eg(m,w) \cdot \gamma^{w-m}(1-\gamma)^{2m-w}
\label{eq:tauG}
\\
p(G^{(\gamma)}_m, \bF_2^w)&=\binom{m}{w-m}\cdot \gamma^{w-m}(1-\gamma)^{2m-w},
\label{eq:pG}
\end{align}
where $\Eg(m,w)$ is the sum of maximal g-embedding numbers 
\begin{align}
\Eg(m,w)=\sum_{y_1^w} \max_{x_1^m} \gbinom{y_1^w}{x_1^m}.
\label{eq:E2wm}
\end{align}
For example, 
\begin{gather}
\Eg(2,3)=16,\ \Eg(m,m)=2^m,\ \Eg(m,2m)=2^{2m}.
\label{eq:e2example}
\end{gather}

Computation of $\Eg(m,w)$ is summarized in Algorithm~\ref{alg:e12} with the input $t=\text{g}$.
Computing $\Eg(m,w)$ requires running over all $x_1^m\in \bF^m$, and running over all the $\binom{m}{w-m}$ possible sets of the inserted positions.
Furthermore, for each such $x_1^m$ and the set of the inserted positions, we run over all $2^w$ possible values of the inserted bits.
For each obtained pair $(x_1^m,y_1^w)$, we count the number of possible insertion configurations (i.e., the set of the inserted positions together with the value of the inserted bits).
For each obtained $y_1^w$, we are storing the current maximum value of $\gbinom{y_1^w}{x_1^m}$ over all $x_1^m$ that have already been processed.
While running over all $x_1^m$, we update the maximum of the g-embedding numbers when we find $x_1^m$, for which $\gbinom{y_1^w}{x_1^m}$ is larger than the current maximum.


\subsection{A Discussion on Computing the Embedding Numbers}

As shown in the numerical results, the obtained bound is not tight.
In this section, we discuss an open combinatorial problem, which can tighten the proposed converse bound for the deletion channel.

One reason for the bound to be loose is that when we bound the code size for $N=nm$, $n>1$, we are providing extra side information to the receiver.
One can decrease the amount of side information or avoid it at all via computing the d-embedding number for large $m$, but this seems to be a very hard problem.
However, to improve the converse bound it is sufficient to obtain an upper bound\footnote{A bound on the d-embedding number is a computational procedure that computes $E_\text{d}(m,w)$ with asymptotic complexity $o(m \cdot \binom{m}{w}\cdot 2^m)$.} on $E_\text{d}(m,w)$, given by \eqref{eq:Emw}, for larger $m$ (and for all $w=0\dots m$).
We could then use it for $N=nm$, where $n$ is very small (or even $n=1$), thereby providing the receiver with less side information.
If the bounds on $E_\text{d}(m,w)$ are sufficiently tight, the gap to the exact value could be compensated by the decrease of the side information.
We have investigated different methods of obtaining upper bounds on $E_\text{d}(m,w)$, however, the results we obtained were not sufficiently tight. Development of tight (and easily computable) upper bounds will help tighten the converse bounds on the deletion and insertion channels with finite block lengths.

As a side note, we highlight that developing a tight lower bound is not difficult --- one just needs to provide a tractable algorithm that outputs $x_1^n$ for a given $y_1^w$, such that $\binom{x_1^m}{y_1^w}$ is large. 
For example, one can insert symbols one-by-one, extending runs\footnote{A run of $y_1^w$ is a maximum (by inclusion) subvector which consists of equal digits. For example, string $011100$ consists of $3$ runs, $0$, $111$ and $00$. If we want to extend $y_1^w$ with two symbols, we can first choose to extend the second run, which leads to the d-embedding number $\binom{4}{3}$, and then the third run, leading to $\binom{4}{3}\cdot \binom{3}{2}=12$. Further extending the second run would lead to multiplying the existing d-embedding number by $\binom{5}{3}/\binom{4}{3}=2.5$ instead of $\binom{3}{2}$. Observe that the number obtained in this way is in fact a lower bound on the true d-embedding number of the obtained strings, and, subsequently, a lower bound on the true maximum d-embedding number $\displaystyle\max_{x_1^m}\binom{x_1^m}{y_1^w}_\text{d}$.} of $y$ which lead to the maximum d-embedding number.

An upper bound, however, is much harder.
Intuitively, the d-embedding number comprises contributions of two simple types, that are interconnected in a highly sophisticated way.
The contribution of the first type is distributing digits of a run of $y_1^w$ over a run of $x_1^m$.
For example, $\binom{0000111}{011}_\text{d}=\binom{4}{1}\cdot\binom{3}{2}$.
If there is one-to-one correspondence between runs of $y_1^w$ and $x_1^m$, the d-embedding number is a product of the binomial coefficients.
The second contribution is distributing runs of $y_1^w$ over runs of $x_1^m$.
For example, in $\binom{101010}{10}_\text{d}=6$ there is no contribution of the first type, and the combinations are generated solely by distributing the runs themselves.
In $\binom{110100}{10}_\text{d}=9$ there are contributions of both types.
In most cases there are both types of contributions, complicating the calculation of tight upper bounds.



\section{Other Bounds and Approximations}
\label{s:other}

\subsection{The Greedy Achievability Bound}
\label{s:gavb}

In this section, we present an algorithm to compute an achievability bound for a general channel.
The complexity of the algorithm scales at least linearly in the code size, so for the case of binary input channels, the complexity is at least exponential in the input length.
We primarily use this algorithm to compare our converse bound with a tight achievability bound.

Consider a general channel $W:\mX\tosq\mY$ and a code $\mC\subseteq \mX$ of size $M$.
We assume that the transmitted codewords are distributed uniformly over $\mC$.
The ML decoder, when observing $y$, outputs $\arg\max_{x\in\mC} W(y|x)$.
Thus, the probability of correct decoding is given by
\begin{align}
P_{\text{corr}}(\mC) &= \frac{\pi(\mC)}{M} 
\label{eq:Pc}
\\
\pi(\mC) &= \sum_{y\in\mY}\max_{x\in\mC} W(y|x)
\label{eq:pi}
\end{align}

\normalsize
We propose to construct the code $\mC$ by adding codewords one by one in a greedy manner, minimizing the ML decoding error probability.
The algorithm is given in Algorithm~\ref{alg:gavb}.
Note that we override the values from $\mX$ and $\mY$ by integers, which can be attached to the elements of the sets by ordering $\mX$ and $\mY$.
The array \var{XD}, indexed by $x$, is equal to $\pi(\mC\cup\set{x})-\pi(\mC)$.
The idea is to add the codeword $x$, which maximizes $\var{XD}[x]$, to the code $\mC$, and update the values of $\var{XD}[x]$.

Note that the tightness of the bound depends on the selection of the codeword in line~\ref{avb:l:getx} among the codewords which maximize $\var{XD}[x]$, if there are a few of them.
In our implementation, in line~\ref{avb:l:getx} function \var{PopRandomElement()} selects one value from vector $\var{DX}[D]$ uniformly at random, returns its value, and removes it from the vector.

For the deletion channel, the initial choice of all-zero $x$ results in a tight bound. Finally, we note that since it is not possible to give side information and split the insertion/deletion channel into smaller subchannels when obtaining an achievability bound, we simply have a single block, i.e., $n=1$.

\begin{algorithm}
\caption{The greedy achievability bound}
\label{alg:gavb}
\DontPrintSemicolon
\KwIn{Channel probabilities $W(y|x)$}
\KwOut{Array $\ve$ of size $|\mX|$, where $\ve[s]$ is an upper bound on the FER of the optimal code of size $s$}
$\var{Code} \gets |\mX|$-length array of \texttt{False}. It says which elements of $\mX$ are in our code\\
$\var{XD}\gets |\mX|$-length vector of $1$'s; $\pi(\mC\cup\set{x})-\pi(\mC)$, defined in \eqref{eq:pi}, will be stored in $\var{XD}[x]$  \\
$\var{DX} \gets $ an empty priority queue with float keys and vector values. The inverse image of \var{XD} array\\
$\var{DX}[1] \gets (1,2,\dots,|\mX|)$\\
$\var{YP}\gets |\mY|$-vector of $0$'s, $\var{YP}[y]=\max_{x\in\mC} W(y|x)$\\
$\tau \gets 0$ \\
$\ve \gets |\mX|$-vector of floats \\
\For{$M\in\set{1,\dots, |\mX|}$}{ 
	$D \gets $ max. key of \var{DX} \\
  $x \gets \var{DX}[D].\var{PopRandomElement()}$ \label{avb:l:getx} \\
	\If{$\var{DX}[D]$ is empty}{remove $D$ from \var{DX}}
	$\var{Code}[x]\gets \var{True}$ \\
  \For{$y: W(y|x) > \var{YP}[y]$}{    
		$\tau \gets \tau + W(y|x)-\var{YP}[y]$ \\
		\For{$x': W(y|x')>0 \wedge \var{Code}[x']$}{
				$w=\min\set{W(y|x), W(y|x')}$ \\
				\If{$w \leq \var{YP}[y]$}{go to the next iteration}
				$D \gets \var{XD}[x']$ \\
				$\Delta = D-(w-\var{YP}[y])$ \\
				\If{the length of $\var{DX}[D]$ is $1$}{remove $D$ from \var{DX}}
                \Else{remove $x'$ from $\var{DX}[D]$}
				$\var{XD}[x']=\Delta$ \\
				add $x'$ to $\var{DX}[\Delta]$
		}
		$\var{YP}[y]\gets W(y|x)$\\
		$\ve[M]\gets \tau/M$ 
  }
}
\textbf{return} $\ve[1\dots M]$
\end{algorithm}

\subsection{The Normal Approximation}
\label{s:napp}

The normal approximation is an asymptotic upper bound on the code size for a given DMC $W^n$.
It is derived in \cite{polyanskiy2010channel} as
\begin{align}
&\log_2\bfM(W^n,\ve) \nonumber\\
&\leq n\cdot\bfI(W)-\sqrt{n\cdot\bfV(W)}\cdot Q^{-1}(\ve)+O(\log n),
\label{eq:napp}
\end{align}
where $\bfI(W)$ and $\bfV(W)$ are the expectation and the variance of mutual information density under the \emph{optimal input distribution}.
The optimal distribution is the one, that 1) maximizes the expectation of mutual information density; 2) among all distributions, that satisfy the first requirement, it should minimize the variance.
The expectation and variance are given by
\begin{align}
\bfI(W)&=\bfI(W,\op),~\bfV(W) = \bfV(W,\op) 
\label{eq:IWVW}
\\
\op&=\argmin_{p \in \argmax_{p'} \bfI(W,p')} \bfV(W,p) 
\label{eq:op}
\end{align}
\begin{align}
\bfI(W,p)&=\sum_{x\in\mX}p(x)\sum_{y\in\mY}W(y|x)\log_2\frac{W(y|x)}{\sum_{x'\in\mX}p(x')W(y|x')}
\label{eq:IWp}
\\
\bfV(W,p)&=\sum_{x\in\mX}p(x)\sum_{y\in\mY}W(y|x)
\nonumber\\
&\times\br{\log_2\frac{W(y|x)}{\sum_{x'\in\mX}p(x')W(y|x')}-\bfI(W,p)}^2,
\label{eq:VWp}
\end{align}
and the summation is performed only over those pairs $(x,y)$, for which $p(x)W(y|x) > 0$.

Note that, although the term $O(\log n)$ is expected to be ``not very large'', omitting the big-O term does not result in a strict bound; it is only an approximation. 

\section{Numerical Examples}
\label{s:num}

In this section, we compute the proposed converse bounds for the deletion channel and the two insertion channel models. In all these cases, computation of the values of $\Ed(m,w)$, $\Ei(m,w)$ and $\Eg(m,w)$ consumes the most computational load, so we explicitly provide these values in the following tables for the largest values of $m$ that we managed to cover.
We do not provide the embedding numbers for small values of $m$, since in this case the embedding numbers can be computed easily and quickly.

We also provide the software implementation of the LO-CVB, the GAVB, and the normal approximation for the cases of deletion, and Gallager insertion channels in \cite{morozov25software}.

\subsection{Bounds for the Deletion Channel}
\label{s:nrd}

\begin{table}
\centering
\small
\begin{tabular}{|@{\hskip.08cm}r@{\hskip.08cm}||@{\hskip.08cm}r@{\hskip.08cm}|@{\hskip.08cm}r@{\hskip.08cm}|@{\hskip.08cm}r@{\hskip.08cm}|@{\hskip.08cm}r@{\hskip.08cm}|@{\hskip.08cm}r@{\hskip.08cm}|}
\hline
$w$& $m=20$   & $m=21$   & $m=22$    & $m=23$    & $m=24$ \\\hline\hline
0  & 1        & 1        & 1         & 1         & 1\\\hline
1  & 40       & 42       & 44        & 46        & 48\\\hline
2  & 580      & 640      & 704       & 770       & 840\\\hline
3  & 5052     & 5894     & 6804      & 7798      & 8912\\\hline
4  & 30932    & 37994    & 46148     & 55508     & 66590\\\hline
5  & 142184   & 184954   & 237180    & 299834    & 375440\\\hline
6  & 514682   & 708084   & 956052    & 1276366   & 1698526\\\hline
7  & 1481532  & 2216868  & 3191242   & 4504570   & 6253452\\\hline
8  & 3671204  & 5690200  & 8624830   & 12874990  & 19252060\\\hline
9  & 7501642  & 12575196 & 20507658  & 32366540  & 49728000\\\hline
10 & 12986826 & 23446602 & 40757978  & 68846108  & 113599680\\\hline
11 & 18226482 & 35815610 & 69815062  & 126499426 & 221132888\\\hline
12 & 24024636 & 49223870 & 98366644  & 192944942 & 375377726\\\hline
13 & 27877130 & 62086746 & 130971724 & 266328578 & 530775466\\\hline
14 & 27614704 & 67882662 & 156701316 & 341932794 & 712042186\\\hline
15 & 23235832 & 63810994 & 162279170 & 387072452 & 873441674\\\hline
16 & 17051216 & 51413026 & 145310300 & 381739786 & 938630462\\\hline
17 & 11135474 & 36780178 & 112905556 & 326856142 & 885064326\\\hline
18 & 6263626  & 23474060 & 79003896  & 246511952 & 727890178\\\hline
19 & 2928320  & 12917734 & 49317072  & 169050912 & 535748298\\\hline
20 & 1048576  & 5933988  & 26587726  & 103291092 & 360429780\\\hline
21 & 0        & 2097152  & 12015100  & 54626852  & 215728518\\\hline
22 & 0        & 0        & 4194304   & 24310736  & 112058162\\\hline
23 & 0        & 0        & 0         & 8388608   & 49157604\\\hline
24 & 0        & 0        & 0         & 0         & 16777216 \\\hline

\end{tabular}
\caption{Complete sets of $\Ed(m,w)$, defined in \eqref{eq:Emw}, for $20\leq m \leq 24$.}
\label{t:emwc}
\end{table}

\begin{table*}
\centering
\footnotesize
\begin{tabular}{|@{\hskip.08cm}r@{\hskip.08cm}||@{\hskip.08cm}r@{\hskip.08cm}|@{\hskip.08cm}r@{\hskip.08cm}|@{\hskip.08cm}r@{\hskip.08cm}|@{\hskip.08cm}r@{\hskip.08cm}|@{\hskip.08cm}r@{\hskip.08cm}|@{\hskip.08cm}r@{\hskip.08cm}|@{\hskip.08cm}r@{\hskip.08cm}|@{\hskip.08cm}r@{\hskip.08cm}|}
\hline
$w$        &$m=25$      & $m=26$     & $m=27$     & $m=28$      & $m=29$      & $m=30$      & $m=31$      & $m=32$     \\\hline\hline
0          & 1          & 1          & 1          & 1           & 1           & 1           & 1           & 1          \\\hline
1          & 50         & 52         & 54         & 56          & 58          & 60          & 62          & 64         \\\hline
2          & 912        & 988        & 1066       & 1148        & 1232        & 1320        & 1410        & 1504       \\\hline
3          & 10104      & 11392      & 12816      & 14328       & 15948       & 17720       & 19590       &            \\\hline
4          & 78972      & 92926      & 108610     & 126674      & 146544      &             &             &            \\\hline
5          & 469008     & 577354     & 704666     & 851968      &             &             &             &             \\\hline
6          & 2213780    & 2852464    & 3632106    &             &             &             &             &             \\\hline
7          & 8530280    & 11466044   &            &             &             &             &             &             \\\hline
8          & 27789992   &            &            &             &             &             &             &             \\\hline
$m\!-\!8$  & 2240637824 &            &            &             &             &             &             &             \\\hline
$m\!-\!7$  & 2025269262 & 4581040100 &            &             &             &             &             &             \\\hline
$m\!-\!6$  & 1608230308 & 3531630622 & 7718500020 &             &             &             &             &             \\\hline
$m\!-\!5$  & 1159842832 & 2502307000 & 5381260108 & 11536866800 &             &             &             &             \\\hline
$m\!-\!4$  & 765854520  & 1622088172 & 3425200596 & 7212131480  & 15145733976 &             &             &             \\\hline
$m\!-\!3$  & 449410072  & 934052936  & 1937255470 & 4010269676  & 8287188072  & 17098402748 & 35227269292 &             \\\hline
$m\!-\!2$  & 229543200  & 469602612  & 959610890  & 1958866402  & 3994845154  & 8139842074  & 16572346438 & 33715641626 \\\hline
$m\!-\!1$  & 99341908   & 200653638  & 405093026  & 817473192   & 1648993508  & 3325101056  & 6702602476  & 13506588908 \\\hline
$m$        & 33554432   & 67108864   & 134217728  & 268435456   & 536870912   & 1073741824  & 2147483648  & 4294967296  \\\hline
\end{tabular}
\caption{Partial sets of $\Ed(m,w)$ for $25 \leq m \leq 32$.}
\label{t:emwp}
\end{table*}

\begin{table}
\centering
\begin{tabular}{|l|l|l|l|l|}
\hline
$n$ &  $m=5$ &  $m=22$ & $m=23$ & BEC, $N=23n$ 
\\ \hline
$1$      & $\underline{0.71688}$ & $\underline{0.55239}$ & $\underline{\textbf{0.54775}}$ & $0.780436$ \\
$2$      & $\underline{0.69929}$ & $\underline{0.58012}$ & $\underline{\textbf{0.57346}}$ & $0.775619$ \\
$4$      & $0.81882$ & $\underline{0.61719}$ & $\underline{\textbf{0.59406}}$ & $ 0.77818$ \\
$8$      & $0.81077$ & $\underline{\textbf{0.62095}}$ & $\underline{0.62193}$ & $0.782655$ \\
$16$     & $0.80675$ & $\underline{\textbf{0.65946}}$ & $\underline{0.66192}$ & $0.786081$ \\
$32$     & $0.80473$ & $\underline{0.66391}$  & $\underline{\textbf{0.66262}}$ & $0.789518$ \\
$64$     & $0.80373$ & $\underline{\textbf{0.70137}}$ & $\underline{0.70186}$ & $0.792199$ \\
$128$    & $0.80323$ & $\underline{\textbf{0.70135}}$ & $\underline{0.70179}$ & $0.794273$ \\
$256$    & $0.80297$ & $0.73575$ & $\underline{\textbf{0.70211}}$ & $0.795865$ \\
$512$    & $0.80285$ & $0.73572$ & $\textbf{0.73417}$ & $ 0.79702$ \\
$1024$   & $0.80279$ & $0.73571$ & $\textbf{0.73416}$ & $0.797867$ \\
$\infty$ & $0.80272$ & $0.73569$ & $0.73414$ & $ 0.8$ \\
\hline                                             
\end{tabular}
\caption{The LO-CVB on code rate for deletion channel with $\delta=0.2$, for target FER $\ve=0.2$. The upper bound on the capacity is $\textbf{0.491}$ \cite{rahmati15upper}. In bold are the best bounds for given $n$. Underlined are the cases when LO-CVB is not equal to the MO-CVB.}
\label{t:cvbs}	
\end{table}

\begin{figure}
\begin{subfigure}{0.5\textwidth}
\includegraphics[width=\textwidth]{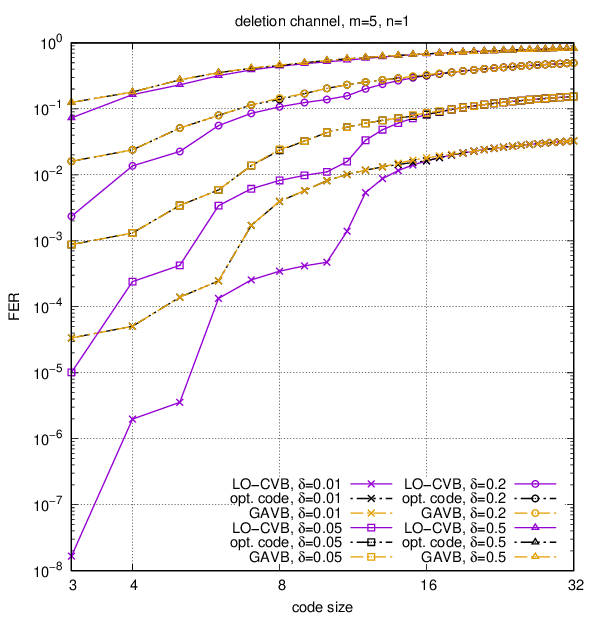}%
\caption{Optimal codes for channel $D_5^{(\delta)}$.}
\label{fig:optdel}
\end{subfigure}
\hfill
\begin{subfigure}{0.5\textwidth}
\includegraphics[width=\textwidth]{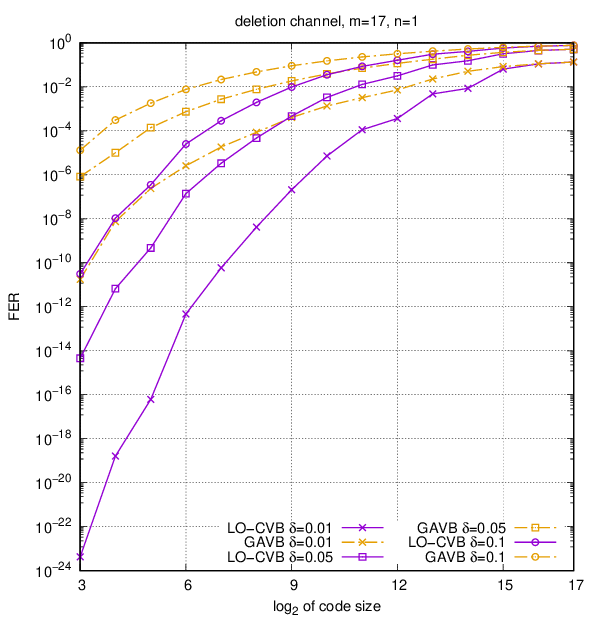}%
\caption{GAVB for channel $D_{17}^{(\delta)}$.}
\label{fig:avbdel}
\end{subfigure}
\caption{The optimal code performance, the GAVB and the LO-CVB for the case of $n=1$.}
\label{fig:n1del}
\end{figure}

\begin{figure*}
\begin{minipage}{0.5\linewidth}
\centering
\begin{subfigure}{\linewidth}
\includegraphics[width=\textwidth]{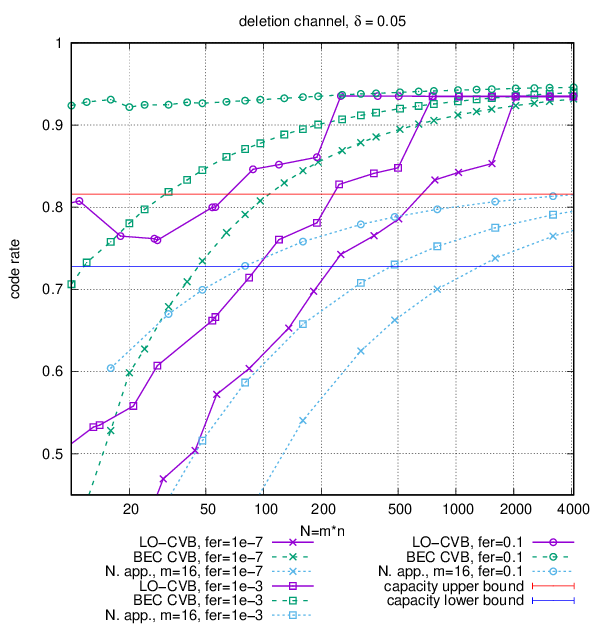}%
\caption{$\delta=0.05$}%
\label{fig:cvbdel-pd-0.05}%
\end{subfigure}
\end{minipage}\hfill
\begin{minipage}{0.5\linewidth}
\centering
\begin{subfigure}{\linewidth}
\includegraphics[width=\textwidth]{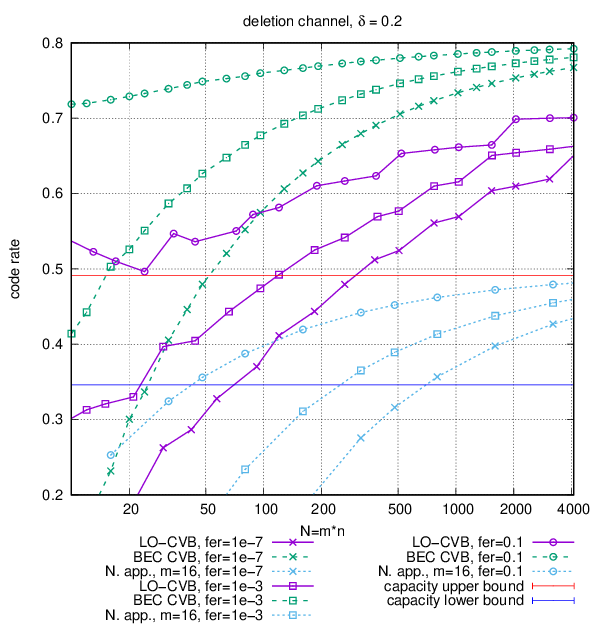}%
\caption{$\delta=0.2$}%
\label{fig:cvbdel-pd-0.2}%
\end{subfigure}
\end{minipage}\hfill
\begin{minipage}{0.5\linewidth}
\centering
\begin{subfigure}{\linewidth}
\includegraphics[width=\linewidth]{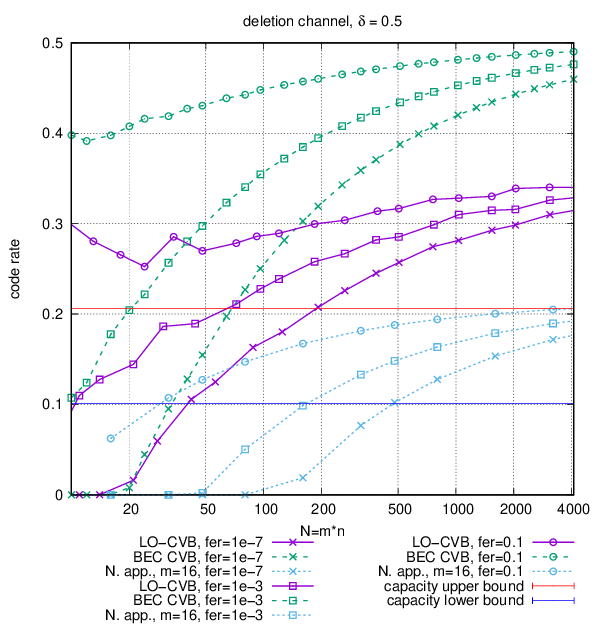}%
\caption{$\delta=0.5$}%
\label{fig:cvbdel-pd-0.5}%
\end{subfigure}
\end{minipage}\hfill
\begin{minipage}{0.5\linewidth}
\centering
\begin{subfigure}{\linewidth}
\includegraphics[width=\linewidth]{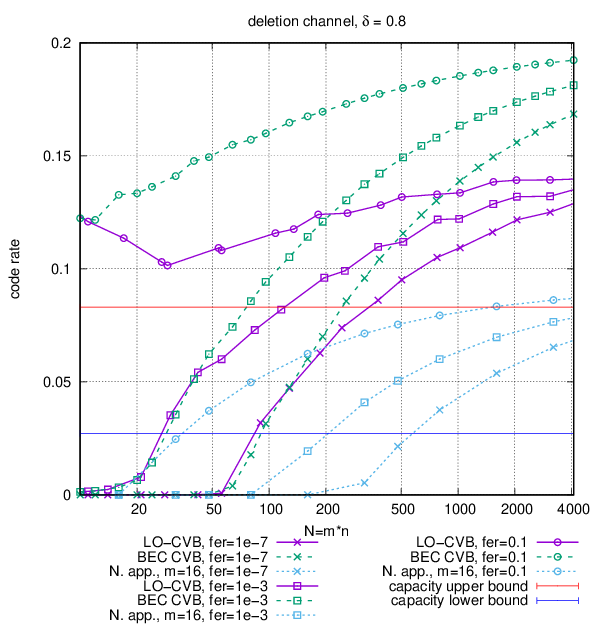}%
\caption{$\delta=0.8$}%
\label{fig:cvbdel-pd-0.8}%
\end{subfigure}
\end{minipage}
\caption{The LO-CVB for the deletion channel $(D_m^{(\delta)})^n$ versus the BEC bound and the normal approximation. The x-coordinate is equal to the total number of input bits $N=mn$.}
\label{fig:locvbdel}
\end{figure*}

In Table~\ref{t:emwc}, we present the complete sets of values of $\Ed(m,w)$, defined in \eqref{eq:Emw}, for $m$ up to $24$.
In Table~\ref{t:emwp}, there are some values of $\Ed(m,w)$ for $m$ up to $32$.
The values in these tables require the most computational effort.
All values of the LO-CVB for the deletion channel below can be obtained by a very simple algorithm of computing \eqref{eq:locvbdel} with complexity\footnote{Which is negligible compared to $O(m\cdot \binom{m}{w} \cdot 2^m)$ operations, needed for computing $\Ed(m,w)$ --- see footnote \ref{ft:opt}.} $O(m\cdot 2^m)$, using only the data from the Tables~\ref{t:emwc}--\ref{t:emwp}.

Using the complete sets of $\Ed(m,w)$ from Table~\ref{t:emwc}, we optimized the values of bound \eqref{eq:Mn} for binary deletion channel with deletion probability $\delta=0.2$ and output FER $\ve=0.2$.
The LO-CVB \eqref{eq:locvbdel} is compared to the BEC bound \eqref{eq:bec} in Table~\ref{t:cvbs}.
Note that, if the complete sets are available, the LO-CVB (underlined in the Table) becomes the MO-CVB (not underlined) for large $n$.

In Fig.~\ref{fig:n1del}, the LO-CVB is compared to the greedy achievability bound (GAVB), presented in Section~\ref{s:gavb}, as well as to the performance of the optimal codes.
The optimal codes are constructed in a similar fashion to the GAVB fashion; namely, we consider all possible codes of length $m=5$ and compute their ML decoding error probability by \eqref{eq:Pc}--\eqref{eq:pi}.
One can see that, at least for $m=5$ (the largest $m$ for which we could construct optimal codes), the GAVB almost coincides with the performance of optimal codes, and the LO-CVB is substantially looser.
For $m=17$ (the largest $m$ for which we could compute the GAVB), the gap between the GAVB and the LO-CVB is large, and, as we expect the GAVB to be very tight, the gap is largely because of looseness of the LO-CVB.
Note, however, that the greedy GAVB is only available for $n=1$.
The main problem of generalizing the GAVB to a larger input length is that any AVB for an improved channel is not an AVB for an original channel, so the standard trick of replacing $D_{mn}^{(\delta)}$ with $(D_{m}^{(\delta)})^n$ does not work, contrary to the case of CVB.

In Fig.~\ref{fig:locvbdel}, the LO-CVB \eqref{eq:locvbdel}, minimized over $20\leq m \leq 32$ using both complete and partial sets of $\Ed(m,w)$ from Tables~\ref{t:emwc}--\ref{t:emwp}, versus the BEC bound \eqref{eq:bec} and the normal approximation \eqref{eq:napp} are presented for $\delta\in\set{0.05,0.2,0.5,0.8}$.
Also, the lower and upper bounds \cite{fertonani10novel} on the capacity of the deletion channel are presented. 
One can see that in most cases, the LO-CVB is better than the BEC bound.
Moreover, the LO-CVB becomes the MO-CVB for large $n$, and from Table~\ref{t:cvbs} it can be seen that the MO-CVB is almost independent of $n$, and the BEC bound grows with $n$ and converges to $1-\delta$, so for larger $n$ the LO-CVB behaves better than the BEC bound.
The results of the LO-CVB are also compared to the lower \cite{fertonani10novel} and upper \cite{rahmati15upper} bounds on the capacity of the deletion channel (see the horizontal lines).
For lengths up to $200$ the LO-CVB is lower than the upper bound on the capacity.
For small lengths, the LO-CVB sometimes is even lower than the \emph{lower} bound on the capacity.
Note that the capacity does not serve as a lower or upper bound on the best achievable rate for any finite length and frame error probability.
The optimal distribution for the normal approximation is obtained using the Blahut-Arimoto Algorithm.
The normal approximation for most cases results in a lower rate than the LO-CVB. Recall, however, that this is only an approximation.

We optimized the bound over $m$ for fixed values of the total number of bits $N=m\cdot n$, allowing the resulting $mn$ to vary in a $\pm 15\%$ gap for each point. We observe a non-monotonic behavior of the code rates as a function of the block lengths. This is caused by the fact that for some specific values of $m$, $N$ and FER, the layers sometimes work better than for the neighboring $N$ values.
Moreover, in the region of very small $N$ the values of $m$ are also very small.
For a smaller $m$, the bound is looser, which makes the bound decrease with $N$, when $N$ is small; although with larger $N$ the maximum achievable code rate is expected to increase.
The discrete steps in the change of the bound with $N$ are caused by the switching of the optimal layers.

The described above effects take place in the LO-CVB for the other channel models as well; similar explanations apply for the results given for the other channel models below.

\subsection{Bounds for the Insertion Channel}
\label{s:nri}

\begin{table}
\centering
\footnotesize
\begin{tabular}{|@{\hskip.08cm}r@{\hskip.08cm}||@{\hskip.08cm}r@{\hskip.08cm}|@{\hskip.08cm}r@{\hskip.08cm}|@{\hskip.08cm}r@{\hskip.08cm}|@{\hskip.08cm}r@{\hskip.08cm}|@{\hskip.08cm}r@{\hskip.08cm}|}
\hline
$w-m$ & $m=12$   & $m=13$    & $m=14$     & $m=15$     & $m=16$      \\\hline\hline
0     & 4096     & 8192      & 16384      & 32768      & 65536       \\\hline
1     & 32144    & 66176     & 135848     & 278208     & 568596      \\\hline
2     & 177116   & 379004    & 805620     & 1702912    & 3582504     \\\hline
3     & 759508   & 1707676   & 3788808    & 8318504    & 18109988    \\\hline
4     & 2586072  & 6213556   & 14602676   & 33694776   & 76584260    \\\hline
5     & 7086760  & 18360020  & 46192552   & 113374076  & 272371584   \\\hline
6     & 15949704 & 44977192  & 121838048  & 319524944  & 815808212   \\\hline
7     & 29385408 & 91812020  & 271333116  & 767085072  & 2091803552  \\\hline
8     & 43743196 & 154764488 & 507547472  & 1568497212 & 4621034516  \\\hline
9     & 51645708 & 212750832 & 790506648  & 2712926188 & 8745295672  \\\hline
10    & 47421160 & 234380952 & 1012050660 & 3935486484 & 14095929836 \\\hline
11    & 32915456 & 203590592 & 1048878624 & 4726630364 & 19171214512 \\\hline
12    & 16777216 & 135528448 & 867617752  & 4640216192 & 21738893936 \\\hline
13    & 0        & 67108864  & 556433408  & 3674853240 & 20332857728 \\\hline
14    & 0        & 0         & 268435456  & 2279079936 & 15485164136 \\\hline
15    & 0        & 0         & 0          & 1073741824 & 9315876864  \\\hline
16    & 0        & 0         & 0          & 0          & 4294967296  \\\hline
\end{tabular}
\caption{Complete sets of $\Ei(w,m)$ for $12 \leq m \leq 16$.}
\label{t:e1wmc}
\end{table}

\begin{table*}
\centering
\footnotesize
\begin{tabular}{|@{\hskip.08cm}r@{\hskip.08cm}||@{\hskip.08cm}r@{\hskip.08cm}|@{\hskip.08cm}r@{\hskip.08cm}|@{\hskip.08cm}r@{\hskip.08cm}|@{\hskip.08cm}r@{\hskip.08cm}|@{\hskip.08cm}r@{\hskip.08cm}|@{\hskip.08cm}r@{\hskip.08cm}|@{\hskip.08cm}r@{\hskip.08cm}|@{\hskip.08cm}r@{\hskip.08cm}|@{\hskip.08cm}r@{\hskip.08cm}|}
\hline
$w-m$     & $m=17$	    & $m=18$      & $m=19$      & $m=20$      & $m=21$	    & $m=22$      & $m=23$      & $m=24$      & $m=25$      \\\hline\hline
0         & 131072      & 262144      & 524288      & 1048576     & 2097152     & 4194304     & 8388608     & 16777216    & 33554432    \\\hline
1         & 1160068     & 2363256     & 4808064     & 9770824     & 19835896    & 40232864    & 81537992    & 165129384   & 334198412   \\\hline
2         & 7505776     & 15669256    & 32608844    & 67673416    & 140098772   & 289402524   & 596655900   & 1227968548  & 2523297548  \\\hline
3         & 39152708    & 84146640    & 179921796   & 382968184   & 811859664   & 1714792000  & 3609968460  & 7576850164  & 15859355896 \\\hline
4         & 171932896   & 382112920   & 842184936   & 1843307984  & 4010620052  & 8681299640  & 18705570280 & 40138809072 & 85806348556 \\\hline
5         & 642342604   & 1490975024  & 3414415232  & 7730789144  & 17337105320 & 38567784692 & 85211252364 &             &             \\\hline
6         & 2035759252  & 4978548524  & 11958029908 & 28266487604 &             &             &             &             &             \\\hline
7         & 5537658308  & 14301775216 & 36160689020 &             &             &             &             &             &             \\\hline
8         & 13089265868 &             &             &             &             &             &             &             &             \\\hline
\end{tabular}
\caption{Some values of $\Ei(w,m)$ for $17 \leq m \leq 25$.}
\label{t:e1wmp}
\end{table*}

\begin{figure}
\centering
\includegraphics[width=0.48\textwidth]{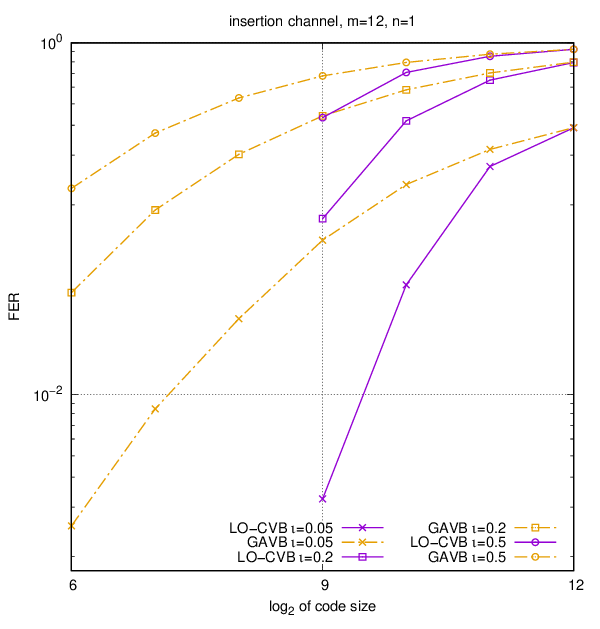}%
\caption{GAVB for channel $I_{12}^{(\iota)}$.}
\label{fig:avbins}
\end{figure}

\begin{figure*}
\begin{subfigure}{0.48\textwidth}
\includegraphics[width=\textwidth]{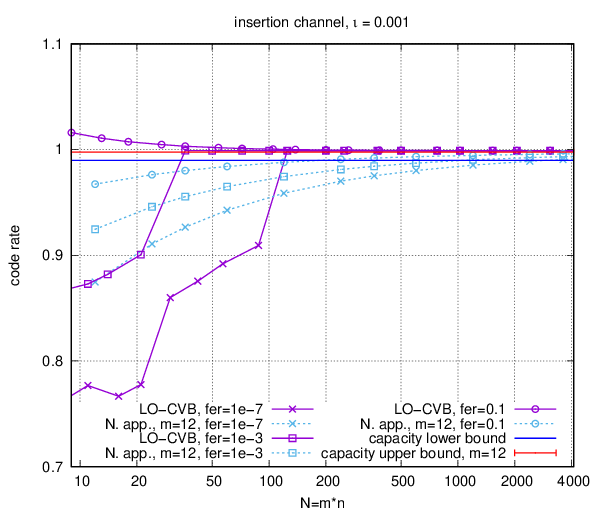}%
\caption{$\iota=0.001$}%
\label{fig:cvbins-pi-0.001}%
\end{subfigure}
\hfill
\begin{subfigure}{0.48\textwidth}
\includegraphics[width=\textwidth]{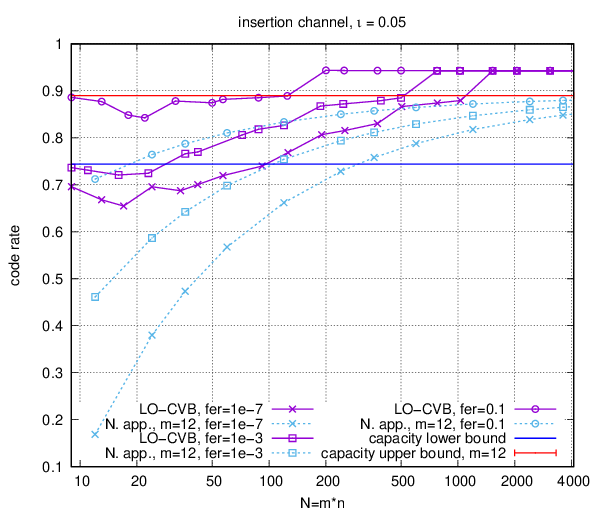}%
\caption{$\iota=0.05$}%
\label{fig:cvbins-pi-0.05}%
\end{subfigure}
\hfill
\begin{subfigure}{0.48\textwidth}
\includegraphics[width=\textwidth]{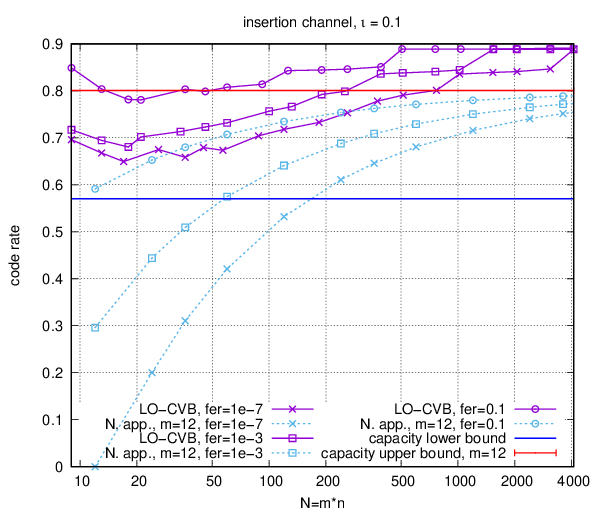}%
\caption{$\iota=0.1$}%
\label{fig:cvbins-pi-0.1}%
\end{subfigure}
\hfill
\begin{subfigure}{0.48\textwidth}
\includegraphics[width=\textwidth]{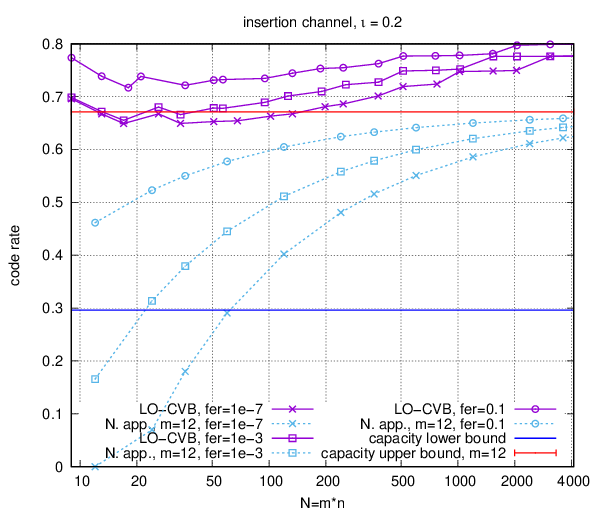}%
\caption{$\iota=0.2$}%
\label{fig:cvbins-pi-0.2}%
\end{subfigure}
\caption{The LO-CVB for the insertion channel $(I_m^{(\iota)})^n$ versus the normal approximation and the lower capacity bound \cite{rahmati13bounds}. The x-coordinate is equal to the total number of input bits $N=mn$.}
\label{fig:locvbins}
\end{figure*}

All the values of i-embedding numbers $\Ei(w,m)$ \eqref{eq:E1wm}, which we managed to compute, are given in Table~\ref{t:e1wmc} (complete sets for $m$ up to $16$) and Table~\ref{t:e1wmp} (partial sets for $m$ up to $25$).
Similarly to the case of the deletion channel, all results for LO-CVB for the insertion channel can be easily computed, using only the provided combinatorial numbers.

In Fig.~\ref{fig:avbins}, the GAVB, computed by Algorithm~\ref {alg:gavb}, is compared to the LO-CVB for the case of $m=12$.
The LO-CVB is larger than zero only for $\log_2M\geq 9$.
As in the deletion channel, the gap between the AVB and the LO-CVB is large.

In Fig.~\ref{fig:locvbins}, the LO-CVB for the insertion channels with $\iota\in\set{0.001,0.05,0.1,0.2}$ is compared to the normal approximation and lower bound on the capacity of the insertion channel from \cite{rahmati13bounds}, Table~II, and to the upper bound on the capacity, which was obtained by the BAA algorithm for $m=12$.
For very low $\iota$ and short length, the LO-CVB is even lower than the normal approximation.
The larger the value of $\iota$, the larger the gap between the LO-CVB and the normal approximation, as well as between the normal approximation and the lower bound on the capacity.

\subsection{Bounds for the Gallager Insertion Channel}
\label{s:nrg}

\begin{table}
\centering
\footnotesize
\begin{tabular}{|@{\hskip.04cm}r@{\hskip.04cm}||@{\hskip.04cm}r@{\hskip.04cm}|@{\hskip.04cm}r@{\hskip.04cm}|@{\hskip.04cm}r@{\hskip.04cm}|@{\hskip.04cm}r@{\hskip.04cm}|@{\hskip.04cm}r@{\hskip.04cm}|@{\hskip.04cm}r@{\hskip.04cm}|}
\hline
$w\!-\!m$  & $m=9$    & $m=10$    & $m=11$    & $m=12$       & $m=13$     & $m=14$     \\\hline\hline
$0$    & 512      & 1024      & 2048      & 4096         & 8192       & 16384      \\\hline
$1$    & 4448     & 9240	  & 19096     & 39312        & 80672      & 165120     \\\hline
$2$    & 27160	  & 59808     & 129656	  & 277792       & 590120     & 1245384    \\\hline
$3$    & 121960	  & 291864    & 678704	  & 1545208      & 3458088    & 7629064    \\\hline
$4$    & 403072	  & 1086040   & 2778864	  & 6838496	     & 16342312   & 38175624   \\\hline
$5$    & 980832   & 3049736   & 8778968	  & 23977640	 & 62549688   & 157309776  \\\hline
$6$    & 1651032  & 6342720   & 21480536  & 66204408	 & 191466944  & 528065728  \\\hline
$7$    & 1835008  & 9320016   & 38892312  & 142511256	 & 470847240  & 1441336312 \\\hline
$8$    & 1179648  & 9175040   & 50675392  & 229081944	 & 900919744  & 3181097000 \\\hline
$9$    & 262144	  & 5242880	  & 44564480  & 267959360	 & 1304833040 &            \\\hline
$10$   & 0        & 1048576	  & 23068672  & 213909504	 & 1383927488 &            \\\hline
$11$   & 0        & 0         & 4194304	  & 100663296    & 1006632960 & 	       \\\hline
$12$   & 0        & 0         & 0         & 16777216     & 436207616  &            \\\hline
$13$   & 0        & 0         & 0         & 0            & 67108864	  &            \\\hline
\end{tabular}
\caption{Complete sets of $\Eg(w,m)$ for $9 \leq m \leq 13$ and incomplete set of $\Eg(w, 14)$.}
\label{t:e2wmc}
\end{table}

\begin{figure*}
\begin{subfigure}{0.48\textwidth}
\includegraphics[width=\textwidth]{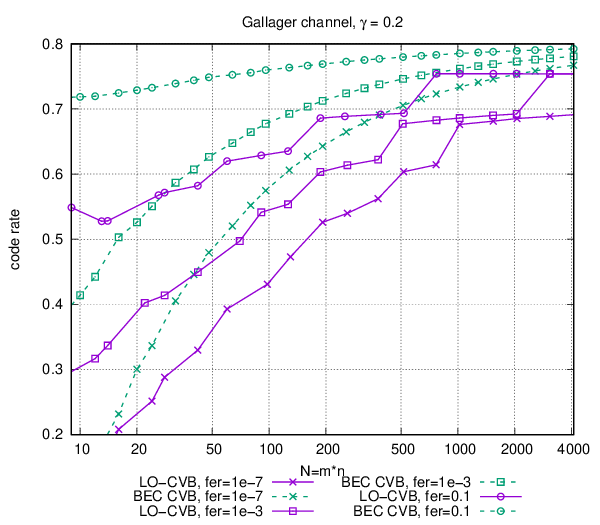}%
\caption{$\gamma=0.2$}%
\label{fig:cvbgal-pg-0.2}%
\end{subfigure}
\hfill
\begin{subfigure}{0.48\textwidth}
\includegraphics[width=\textwidth]{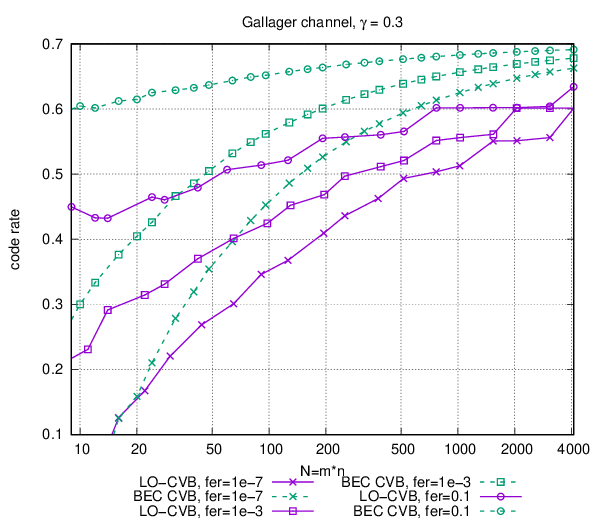}%
\caption{$\gamma=0.3$}%
\label{fig:cvbgal-pg-0.3}%
\end{subfigure}
\hfill
\begin{subfigure}{0.48\textwidth}
\includegraphics[width=\textwidth]{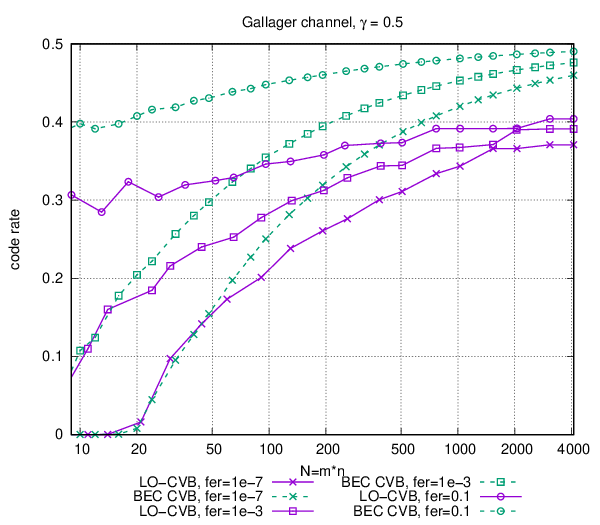}%
\caption{$\gamma=0.5$}%
\label{fig:cvbgal-pg-0.5}%
\end{subfigure}
\hfill
\begin{subfigure}{0.48\textwidth}
\includegraphics[width=\textwidth]{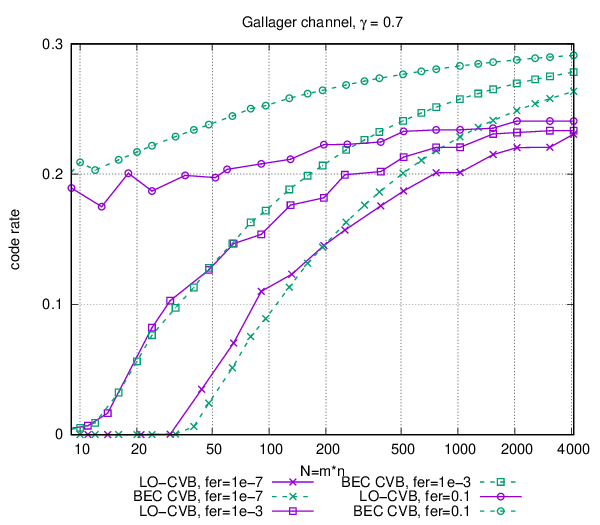}%
\caption{$\gamma=0.7$}%
\label{fig:cvbgal-p-0.7}%
\end{subfigure}
\caption{The LO-CVB for the Gallager's insertion channel $(G_m^{(\gamma)})^n$ versus the BEC bound.
The x-coordinate is equal to the total number of input bits $N=mn$.}
\label{fig:locvbgal}
\end{figure*}

All the values of g-embedding numbers $\Eg(w,m)$ \eqref{eq:E2wm}, which we managed to compute, are given in Table~\ref{t:e2wmc} (complete sets for $m\leq 13$ and partial set for $m=14$).
Similarly to the case of the deletion channel, all results for LO-CVB for the Gallager's insertion channel can be easily computed, using only the provided combinatorial numbers.

In Fig.~\ref{fig:locvbgal}, the LO-CVB for the Gallager's insertion channels with $\gamma\in\set{0.2,0.3,0.5,0.7}$ is compared to the BEC bound.
For the values of $\gamma$ which are closer to $0$ or $1$ and short lengths, the LO-CVB is looser than the BEC bound. For middle values of $\gamma$ and larger lengths, the LO-CVB is tighter.

\subsection{Possible Extensions}

The proposed bounds can be generalized to other channel models. For instance, there are different insertion channel models, including duplication channels, to which the same ideas and approaches would apply. As a further extension, insertions/deletions with memory can also be considered, for instance, employing the model in \cite{morozov24capacity}. In this case, the side information to be provided would include the starting and ending states of the Markov chain describing the insertion/deletion channel, along with the number of received bits in each block. With this side information, the new channel will have a product form (again, with fewer input and output bits), and the same ideas can be applied in a similar way.

\section{Conclusions}
\label{s:conc}
In this paper, we develop upper bounds on the code size for the deletion (insertion) channel with a given deletion (insertion) probability, input length, and target codeword error probability. 
This is done by providing a reference output distribution for the general converse bound from \cite{tan15asymptotic}, which we call a layer-oriented distribution.
The layer-oriented distribution leads to a converse bound tighter than the BEC bound, which is the only existing alternative, for both deletion and insertion channels. We also provide an algorithm for computing a simple achievability bound for a general discrete channel, and apply it to the channels with synchronization errors. The developed converse bounds are obtained by significantly relaxing the symbol-wise converse bounds; however, they remain useful for deletion and insertion channels. 
This is because they are coupled with some side information to make them tractable, and the amount of side information given is less compared to the alternative of the BEC bound. Finally, we note that although the bounds are tighter than existing results, the gap between the converse and achievability bounds remains significant, indicating that there is room for further improvement.

\bibliographystyle{IEEEtran}
\bibliography{coding}

\begin{IEEEbiography}[{{\includegraphics[scale=0.4]{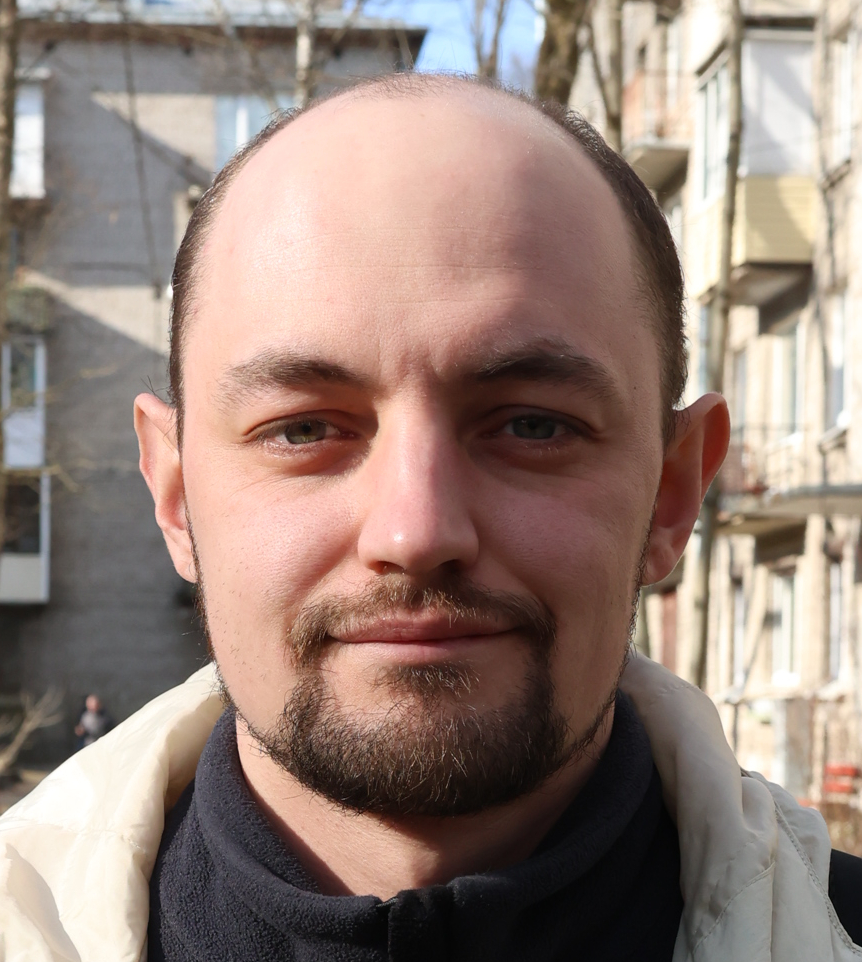}}}]
{Ruslan Morozov} was born in Saint Petersburg, USSR. He received BSc and MSc in Saint Petersburg Polytechnic University, and PhD in Information Theory in ITMO University, the advisor was Peter Trifonov. He has also worked in Huawei Moscow Research Center in the Coding Lab, and in CTAR Lab, Bilkent University, Ankara, under supervision of Tolga Duman. His research interests include coding theory and information theory, specifically polar and LDPC/GLDPC codes, as well as their applications. He is also interested in the combinatorics of the deletion channel.
\end{IEEEbiography}

\begin{IEEEbiography}[{{\includegraphics[scale=0.2]{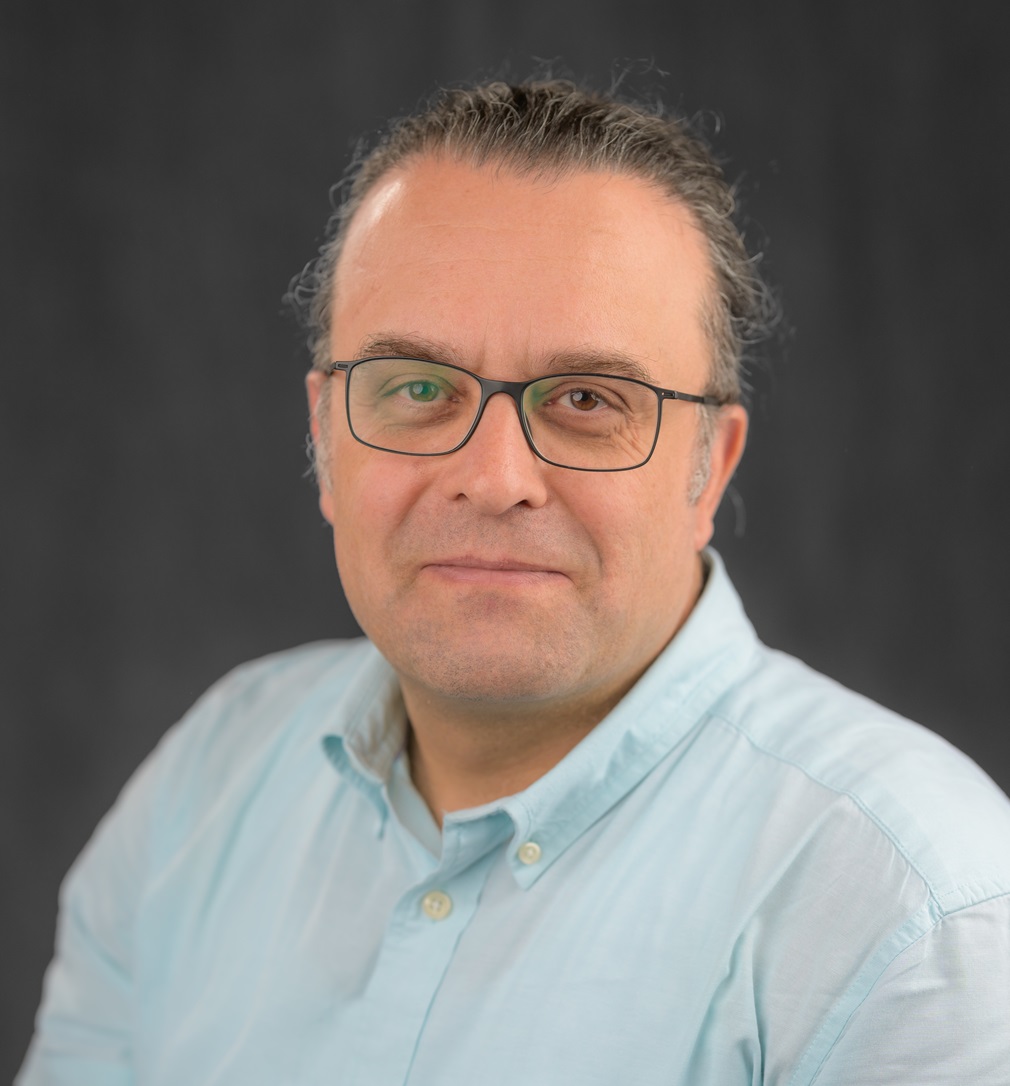}}}]
{Tolga M. Duman} (S'95--M'98--SM'03--F'11) is a Professor
of the Electrical and Electronics Engineering Department at Bilkent
University in Turkey. He received a B.S. degree from Bilkent University,
Ankara, Turkey, in 1993, and the M.S. and Ph.D. degrees from Northeastern
University, Boston, MA, USA, in 1995 and 1998, respectively, all in
electrical engineering. Before joining Bilkent University in September
2012, he was a full professor with the School of ECEE at Arizona State
University. Dr. Duman's current research interests are in systems,
with a particular focus on communications and signal processing, including
wireless and mobile communications, channel coding/modulation, coding
for wireless communications, information theory, and data storage systems,
as well as machine learning for communications and semantic communications/signal
processing.\\
Dr. Duman is a Fellow of the IEEE, a recipient of the National Science
Foundation CAREER Award, the IEEE Third Millennium medal, and a European
Research Council (ERC) Advanced Grant. He served on the editorial
boards of various journals, including IEEE TRANSACTIONS ON WIRELESS
COMMUNICATIONS AND IEEE COMMUNICATIONS SURVEYS AND TUTORIALS. He also
served as Editor-in-Chief of Elsevier's Physical Communication journal
and IEEE TRANSACTIONS ON COMMUNICATIONS. 
\end{IEEEbiography}

\end{document}

%% file: macro.tex
\makeatletter
\let\save@mathaccent\mathaccent
\newcommand*\if@single[3]{%
  \setbox0\hbox{${\mathaccent"0362{#1}}^H$}%
  \setbox2\hbox{${\mathaccent"0362{\kern0pt#1}}^H$}%
  \ifdim\ht0=\ht2 #3\else #2\fi
  }
\newcommand*\rel@kern[1]{\kern#1\dimexpr\macc@kerna}
\newcommand*\wideaccent[2]{\@ifnextchar^{{\wide@accent{#1}{#2}{0}}}{\wide@accent{#1}{#2}{1}}}
\newcommand*\wide@accent[3]{\if@single{#2}{\wide@accent@{#1}{#2}{#3}{1}}{\wide@accent@{#1}{#2}{#3}{2}}}
\newcommand*\wide@accent@[4]{%
  \begingroup
  \def\mathaccent##1##2{%
    \let\mathaccent\save@mathaccent
    \if#42 \let\macc@nucleus\first@char \fi
    \setbox\z@\hbox{$\macc@style{\macc@nucleus}_{}$}%
    \setbox\tw@\hbox{$\macc@style{\macc@nucleus}{}_{}$}%
    \dimen@\wd\tw@
    \advance\dimen@-\wd\z@
    \divide\dimen@ 3
    \@tempdima\wd\tw@
    \advance\@tempdima-\scriptspace
    \divide\@tempdima 10
    \advance\dimen@-\@tempdima
    \ifdim\dimen@>\z@ \dimen@0pt\fi
    \rel@kern{0.6}\kern-\dimen@
    \if#41
      #1{\rel@kern{-0.6}\kern\dimen@\macc@nucleus\rel@kern{0.4}\kern\dimen@}%
      \advance\dimen@0.4\dimexpr\macc@kerna
      \let\final@kern#3%
      \ifdim\dimen@<\z@ \let\final@kern1\fi
      \if\final@kern1 \kern-\dimen@\fi
    \else
      #1{\rel@kern{-0.6}\kern\dimen@#2}%
    \fi
  }%
  \macc@depth\@ne
  \let\math@bgroup\@empty \let\math@egroup\macc@set@skewchar
  \mathsurround\z@ \frozen@everymath{\mathgroup\macc@group\relax}%
  \macc@set@skewchar\relax
  \let\mathaccentV\macc@nested@a
  \if#41
    \macc@nested@a\relax111{#2}%
  \else
    \def\gobble@till@marker##1\endmarker{}%
    \futurelet\first@char\gobble@till@marker#2\endmarker
    \ifcat\noexpand\first@char A\else
      \def\first@char{}%
    \fi
    \macc@nested@a\relax111{\first@char}%
  \fi
  \endgroup
}
\makeatother

\newcommand{\op}{{\overline{p}}}

\newcommand{\mC}{{\mathcal{C}}}

\newcommand{\bfD}{{\mathbf{D}}}

\newcommand{\bF}{{\mathbb{F}}}

\newcommand{\bfI}{{\mathbf{I}}}

\newcommand{\mL}{{\mathcal{L}}}

\newcommand{\oM}{{\overline{M}}}

\newcommand{\bfM}{{\mathbf{M}}}

\newcommand{\obfM}{{\overline{\mathbf{M}}}}

\newcommand{\bN}{{\mathbb{N}}}

\newcommand{\tQ}{{\widetilde{Q}}}

\newcommand{\mS}{{\mathcal{S}}}

\newcommand{\mT}{{\mathcal{T}}}

\newcommand{\bfV}{{\mathbf{V}}}

\newcommand{\mX}{{\mathcal{X}}}

\newcommand{\mY}{{\mathcal{Y}}}

\newcommand{\set}[1]{{\left\{{#1}\right\}}}
\newcommand{\br}[1]{{\left({#1}\right)}}
\newcommand{\brsq}[1]{{\left[{#1}\right]}}

\newcommand{\brabs}[1]{{\left|{#1}\right|}}

\newcommand{\floor}[1]{\lfloor{{#1}\rfloor}}
\newcommand{\defeq}{{\overset{\triangle}{=}}}

\newcommand{\ve}{\varepsilon}
\newcommand{\vp}{\varphi}
\newcommand{\tosq}{\rightsquigarrow}

\newcommand{\var}[1]{\texttt{#1}}

\newtheorem{proposition}{Proposition}